\documentclass[iop,apj,numberedappendix]{emulateapj}
\usepackage{amsmath,amssymb,amstext}
\usepackage[breaklinks,colorlinks,citecolor=blue,linkcolor=magenta]{hyperref} 
\usepackage{graphicx}

\usepackage[all]{hypcap} %Links go to figures; breaks on deluxetables (use \capstartfalse \capstarttrue to fix it)

\usepackage{aas_macros} 
\usepackage{natbib}
\shorttitle{Systematic SN~Ia Nebular Modeling}
\shortauthors{Boty\'{a}nszki \& Kasen}

\begin{document}

\newcommand{\fsix}[1]{$^{56}$#1}
\newcommand{\Msun}[0]{$\rm M_\odot$}
\newcommand{\sub}[2]{${#1}\rm _{#2}$}
\newcommand\T{\rule{0pt}{2.6ex}}       % Top strut
\newcommand\B{\rule[-1.2ex]{0pt}{0pt}} % Bottom strut
\newcommand{\newpoint}[1]{\smallskip \noindent {\bf #1}:}

\title{How do Type Ia Supernova Nebular Spectra Depend on Explosion Properties? Insights from systematic non-LTE modeling}
\author{J\'{a}nos Boty\'{a}nszki\altaffilmark{1}, Daniel Kasen\altaffilmark{1,2,3}}
\affil{\altaffilmark{1}Physics Department, University of California, Berkeley, CA 94720, USA}
\affil{\altaffilmark{2}Astronomy Department and Theoretical Astrophysics Center, University of California, Berkeley, CA 94720, USA}
\affil{\altaffilmark{3}Nuclear Science Division, Lawrence Berkeley National Laboratory, Berkeley, CA 94720, USA}

\begin{abstract}

We present a radiative transfer code to model the nebular phase spectra of supernovae (SNe) in  non-local thermodynamic equilibrium (NLTE).
We apply it to a systematic study of Type~Ia SNe using parameterized 1D models and show how nebular spectral features depend on key physical parameters, such as the time since explosion, total ejecta mass, kinetic energy, radial density profile, and the masses of \fsix{Ni}, intermediate mass elements (IMEs), and stable iron-group elements (IGEs). We also quantify the impact of uncertainties in atomic data inputs. Among the results of the study are: 
(1) The main features of SNe~Ia nebular spectra are relatively insensitive to most physical parameters. Degeneracy among parameters precludes a unique determination of the ejecta properties from spectral fitting. In particular, features can be equally well fit with generic Chandrasekhar mass (\sub{M}{ch}), sub-\sub{M}{Ch}, and super-\sub{M}{Ch} models; 
(2) A sizable ($\gtrsim$0.1~\Msun) central region of stable IGEs, often claimed as evidence for \sub{M}{Ch} models, is not essential to fit the optical spectra and may produce an unusual flat-top [CoIII] profile; 
(3) The strength of [SIII] emission near 9500 \AA\ can provide a useful diagnostic of explosion nucleosynthesis; 
(4) Substantial amounts ($\gtrsim$0.1~\Msun) of unburned C/O mixed throughout the ejecta produce [OIII] emission not seen in observations; 
(5) Shifts in the wavelength of line peaks, sometimes used to infer ejecta geometry, can also arise from line blending effects;
(6) The steepness of the ejecta density profile affects the line shapes, with flatter slopes providing better fits to the  observations of SN~2011fe, offering a constraint on explosion models;
(7) Uncertainties in atomic data affect spectral line ratios by $\sim$30\%, a level similar to the effect of varying physical parameters. 

\end{abstract}

\keywords{supernovae: general --- radiative transfer ---  line: formation --- radiation mechanisms: non-thermal}
\maketitle

\section{Introduction}

Spectra taken of supernovae at late times (in the nebular phase, $\gtrsim$ 100 days after explosion) probe the central regions of the ejecta and thus contain a wealth of information about the explosion, such as nucleosynthetic yields, compositional mixing, and geometry. The quantity and breadth of nebular spectra has grown rapidly in recent years due to international observational efforts. However, further modeling is needed to develop a systematic understanding of how nebular spectra depend on explosion parameters and how atomic data inputs affect spectral modeling. 

Nebular spectra can be used to study the uncertain progenitors of Type Ia supernovae \citep[SNe~Ia,][]{sollerman2004, maeda2010a_asymmetry, mazzali2011, mazzali2012, mazzali2015}. While SNe~Ia are thought to be the result of the explosion of carbon-oxygen white dwarfs (C/O WDs) in a binary system \citep{hoyle1960}, their progenitor systems and explosion mechanisms are still unknown. Despite the success of the empirical width-luminosity relation \citep{phillips1993} to calibrate luminosities of SNe~Ia, systematic variation due to intrinsic SN~Ia diversity is an ongoing challenge for precision cosmology \citep{howell2011}. 

Several candidates for the progenitors of ``normal'' SNe~Ia have been proposed \citep[see][and references therein]{branch1993, wang2012, maoz2014}. In the single-degenerate \sub{M}{Ch} model, the WD gains mass from a non-degenerate binary companion \citep{whelan1973} and ignites carbon burning when the mass approaches $M_{\rm ch} \approx 1.4M_\odot$. In the double-detonation model, a layer of helium gas from a binary companion detonates above the primary C/O WD causing the WD to detonate at a mass \sub{< M}{Ch} \citep{livne1990a_double_detonation, woosley1994, fink2007, guillochon2010, shen2014}. In the double-degenerate model, two WDs coalesce or collide and may detonate violently on impact or subsequent to the merger \citep{iben1984, webbink1984, benz1989, raskin2009, rosswog2009, pakmor2012, kushnir2013, moll2014, kashyep2015}. It may be that SNe~Ia come from most, if not all, of these progenitor channels. A main goal of spectral modeling is to help understand the origin of different events. 

Nebular spectra reveal emission throughout the entire ejecta and so are a valuable probe of density and compositional structure. If the ejecta mass, kinetic energy, and/or compositional yields can be determined, these can differentiate explosion scenarios. For example, it is thought that \sub{M}{Ch} models produce more stable IGEs (such as $^{58}$Ni and $^{54}$Fe) than sub-\sub{M}{Ch} mass models due to burning at higher central densities \citep{iwamoto1999}. Nebular line profiles are also sensitive to global asymmetries and therefore offer a way to study the imprints of the explosion mechanism on ejecta geometry. 

Initial work by \citet{axelrod1980} enabled a number of codes used to model the nebular spectra of supernovae \citep{ruiz-Lapuente1992, liu1997a_nebular, kozma1998a, kozma1998b}. Updated codes use improved atomic data and incorporate more sophisticated radiative transport and non-thermal deposition physics \citep{sollerman2000, mazzali2001, sollerman2004, kozma2005, maeda2006_TypeIc_multidim, mazzali2007b_nebular, maurer2010, dessart2011, jerkstrand2011_1987A, li2012}. Despite the publication of numerous 3D hydrodynamic simulations of SNe~Ia \citep[e.g.,][]{hillebrandt2013}, few such explosion models have been analyzed in the nebular phase \citep[but see][]{kozma2005}. Furthermore, previous nebular modeling work has focused on interpreting events individually \citep{ruiz-Lapuente1992, maeda2010a_asymmetry, maeda2010b_nature, sollerman2004, stehle2005, maurer2011, mazzali2011, mazzali2012, taubenberger2013_oxygen_in_SnIa, mazzali2015, ashall2016}, rather than carrying out parameter studies to identify  the  general dependencies and degeneracies of nebular spectra.

The aim of this work is to systematically study how variations in explosion properties, density and abundance structure, and atomic data inputs affect the spectra of SNe~Ia at late times. To that end, we present a new NLTE code to model the nebular spectra of supernovae. In \S\ref{sec:methods}, we present our method of calculating level populations, non-thermal deposition, temperature and ionization balance, and nebular spectra. In \S\ref{sec:results}, we use a fiducial model to describe the physics of nebular spectral formation in SNe~Ia. We then vary the parameters of the model to probe the sensitivity of the spectra to ejecta mass, composition, and kinetic energy (\S\ref{ssec:t}-\ref{ssec:M_CO}), as well as density profile (\S\ref{sec:density_profiles}) and atomic data inputs (\S\ref{ssec:atomic_study}).

\section{Methods} 
\label{sec:methods}

We have developed a new 3D radiative transfer tool to model nebular spectra of SNe. Given an initial ejecta model, the code calculates the emissivity of each atomic transition by solving for the temperature, ionization state, and NLTE atomic level populations, including non-thermal effects from radioactive decay products, and generates spectra by integrating the radiative transfer equation in a moving medium.

\subsection{Basic Assumptions}
\label{ssec:assumptions}

The underlying supernova ejecta model is specified by the mass density and elemental abundances in each zone on a 3D Cartesian grid.  We assume the ejecta are in homologous expansion (i.e., velocity proportional to radius) which is appropriate for SNe~Ia at a few seconds to days after explosion \citep[e.g.,][]{ropke2005}. The structure of a homologous model at one epoch can be easily scaled to any other time, taking into account compositional changes due to radioactive decay. The models in this paper study SNe~Ia out to 400 days and include the following species: C, O, Si, S, Ca, Fe, Co, and Ni (See Appendix \ref{appx:atomicData} for a description of atomic data sources). 

Our calculations assume stationarity -- i.e., that the gas temperature and level populations reach equilibrium on a timescale short compared to the ejecta expansion timescale. This assumption is reasonable except at rather late times ($\gtrsim $500 days), when thermal and ionization freeze-out may become important \citep{fransson1993,kozma1998a, sollerman2004, fransson2015}. 

Our transport solver assumes that the ejecta are optically thin to radiation. For many supernovae, this is true in the optical and infrared regions at times $\gtrsim 100$~days. However, at blue and ultraviolet wavelengths ($\lesssim 4000$ \AA), the ejecta may remain opaque to iron-group lines for hundreds of days \citep{friesen2017}. This limits the reliability of our nebular models at short wavelengths, an issue that will be addressed in the future by incorporating a more general transport solver.

\subsection{Non-LTE Level Populations}
\label{ssec:nlte}

To calculate spectral emission from the SN nebula, one has to determine the level populations of each ion in the gas. Since LTE is a poor approximation at these epochs,
we solve for each species the set of NLTE equations expressing statistical equilibrium

\begin{equation}
    \label{equ:matrix}
	\mathbf{M}\vec{n} = \bf{0}
\end{equation}
where $\vec{n}$ is a vector of level populations for the species, and the matrix $\mathbf{M}$ encodes the transition rates between the various levels and ionization states \citep[see e.g.,][]{liMcCray1993}. To the set of equations
must be added the  constraint of  number conservation.

A generalized form for the statistical equilibrium rate equations is

\begin{equation}
    \label{equ:rate_equation}
	n_{\rm I,i}\left(R_{\rm I,i} + \sum_{\rm j \neq i}\mathcal{E}_{\rm ij}\right) = \sum_{\rm k \neq i} n_{\rm k}\mathcal{E}_{\rm ki} + n_{\rm I+1,gs}n_{\rm e}\alpha_{\rm i}
\end{equation}
where \sub{n}{I,i}\ is the level population of the i-th level of the ion with I-th ionization state, \sub{n}{e} is the electron density, \sub{R}{I,i}\ is the total ionization rate from the i-th to the I+1-th ionization state (including photoionization, collisional ionization, and non-thermal ionization), and \sub{\alpha}{i}(T) is the total recombination coefficient from ionization state I's ground state to the i-th level (including radiative, dielectronic, and three-body recombination). \sub{\mathcal{E}}{ij} are bound-bound rate coefficients between levels i and j, including spontaneous emission, stimulated absorption/emission, and collisional excitation/de-excitation. 

The microphysics relevant to the nebular phase calculation is encoded into the transition matrix of eq. \ref{equ:matrix}. For the SN~Ia problem, we treat collisional (``electron-impact'') bound-bound and bound-free processes, radiative and dielectronic recombination, and non-thermal deposition of \fsix{Co} energy into heating, excitation, and ionization channels. Following \citet{nahar1997_C_N}, we assume that all collisional ionizations rates are ground state rates, and that radiative recombinations come from the ground state. Details of the implementation of these processes are given in Appendix \ref{appx:implementation}.

\subsection{Non-thermal Effects}
\label{sssec:NT}
Energy deposition due to the radioactive decay chain \fsix{Ni}$\rightarrow$\fsix{Co}$\rightarrow$\fsix{Fe} is the dominant energy source for SNe~Ia in the nebular phase \citep{colgate1969,kuchner1994}.  Since \fsix{Ni} decays on a timescale of $\rm \tau_{Ni} = 7.7$ days, its contribution to heating is minor by nebular times, while \fsix{Co} decay ($\rm \tau_{Co} = 111.3$ days) is generally the most important source of radioactive energy. 
Other radioactive isotopes with half-lives in the nebular range (e.g., $^{22}$Na, $^{35}$S, $^{45}$Ca, $^{46}$Sc, $^{49}$V, $^{54}$Mn, $^{55}$Fe, $^{57}$Co, $^{58}$Co, $^{65}$Zn, and $^{68}$Ge) are usually not produced in enough abundance to significantly contribute to heating at the epochs we consider here \citep{iwamoto1999, seitenzahl2013}, and  are not included in the present calculations.

We determine the radioactive energy deposition rate in each zone using a  3D Monte Carlo radiation transport scheme \citep{kasen2006} that samples gamma-ray wavelengths from the radioactive decay lines and treats energy losses due to Compton scattering and photoionization. Positrons from $^{56}$Co decay are assumed to be trapped locally, in accordance with recent observations of SNe~Ia at late times (\citet{leloudas2009,kerzendorf2014}, but see \citet{dimitriadis2017}).

The gamma-rays from \fsix{Co} decay produce high-energy electrons ($E_0 \sim 1$ MeV) that interact with the ejecta through heating (Coulomb interactions), ionization, and excitation. These interactions affect the temperature and ionization state of the gas. The rates of excitation and ionization, respectively, by non-thermal electrons/positrons are given by

\begin{equation}
\label{eqn:R_ex}
R_{\rm ex} = \frac{\dot{\epsilon}_{\rm rad} \eta_{\rm ij}}{\Delta E_{\rm ij} n_{\rm i}}
\end{equation} 

\begin{equation}
\label{eqn:R_ion}
R_{\rm ion} =  \frac{\dot{\epsilon}_{\rm rad} \eta_{\rm k}}{I_{\rm k} n_{\rm k}}
\end{equation}
where the transition from level $i$ to level $j$ (with $j > i$) has energy \sub{\Delta E}{ij},  \sub{n}{i} is the level population of the i-th atomic level, \sub{I}{k}\ is the ionization potential of the ion indexed by k having number density \sub{n}{k}, \sub{\dot{\epsilon}}{\rm rad} is the radioactive energy deposition rate per unit volume, and \sub{\eta}{ij} and \sub{\eta}{k} refer to the non-thermal excitation and ionization deposition fractions, respectively, described in appendix \ref{appx:nonThermal}.

\subsection{Temperature and Ionization Balance} \label{ssec:tempIonBalance}

We use an iterative non-linear solver to determine the free electron density and temperature in each zone. The electron density is constrained by the charge conservation condition that all free electrons come from ionization, i.e. 

\begin{equation}
    \label{equ:charge}
    n_{\rm e} = \sum_{\rm k,i} n_{\rm k,i}i
\end{equation}
where \sub{n}{k,i} represents the population of the i-th ionization stage of the k-th species (i=0 is neutral).
Given that the non-thermal deposition fractions (\sub{\eta}{ij} and \sub{\eta}{k} in eqs. \ref{eqn:R_ex}-\ref{eqn:R_ion}) are level-population dependent, we recalculate them during each iterative step until convergence is reached. 

The temperature of each zone is calculated from the balance of heating from radioactive decay and cooling due to line emission.  The line emissivity (units ergs~s$^{-1}$~cm$^{-3}$) from a transition between the i-th and j-th levels of an ion ($E_{\rm j} > E_{\rm i}$) is

\begin{equation}
    \label{eqn:epsilon}
	\dot{\epsilon}_{\rm ij}(T) =  h\nu_{\rm ij}A_{\rm ij}n_{\rm j}
\end{equation}
where \sub{\nu}{ij} is the photon frequency of the transition, \sub{A}{ij} is the spontaneous radiative decay rate (Einstein A coefficient), and \sub{n}{j} is the population of the j-th level of the ion. The temperature sensitivity of  $\dot{\epsilon}_{\rm ij}$ arises  primarily from the temperature-dependent collisional excitation rates, which are important in setting the populations $n_{\rm j}$ for low lying states.

The total emissivity per unit volume is then a sum over all line transitions which, in equilibrium, is equal to the energy deposition by radioactivity
\begin{equation}
    \label{eqn:cooling}
	\sum_{\rm i,j} \dot{\epsilon}_{\rm ij} = \dot{\epsilon}_{\rm rad}
\end{equation}
where \{i,j\} runs over all transitions. Note that Equation~\ref{eqn:cooling} represents a balance of all emission and deposition processes, not just the thermal processes.

We use an iterative Brent–Dekker method to solve the non-linear equations \ref{equ:charge} and \ref{eqn:cooling} for the temperature and electron density in each zone. For each iteration, we solve the NLTE rate equations to determine the ionization/excitation state and emissivity. The values and $T$ and $n_e$ are then adjusted and the procedure iterated until thermal and ionization equilibrium is reached. A single zone typically converges within 200-400 iterative steps to reach an accuracy of 0.1\% in temperature. 

\subsection{Spectrum Calculation} 
\label{ssec:spectrumCalc}

Once the line emissivities (eq. \ref{eqn:epsilon}) are known for all transitions in a converged model, we can integrate the emission for any arbitrary geometry to determine the nebular spectrum of the SN. Assuming homologous expansion ($r = vt$) and choosing the direction to the observer to be the $z$-axis, the non-relativistic Doppler effect along the line of sight is

\begin{equation}
    \label{eqn:doppler_z}
	\lambda_{\rm obs} = \lambda_{\rm ij} \left(1 - \frac{z}{ct} \right)
\end{equation}
where $\lambda_{\rm obs}$ is the observed wavelength of a line transition with rest-frame wavelength \sub{\lambda}{ij}. Equation~\ref{eqn:doppler_z} allows us to associate the observed flux at wavelength $\lambda_{\rm obs}$ with the line emission integrated over a specific plane (perpendicular to the $z$-axis) sliced through the ejecta. For isolated transitions, this mapping can be used to constrain the geometrical distribution of material along the line of sight \citep[see, e.g.,][]{shivvers2013}. Since the bulk of the emission typically comes from ejecta with velocities $\lesssim $10,000 km/s, using the non-relativistic Doppler shift leads to errors in z-mapping of only $\lesssim 3\%$.

Assuming the ejecta are optically thin, the observed specific intensity can be expressed as the integral
\begin{equation}
L_\lambda(\lambda) d\lambda = \left[\sum_{\rm i,j}\frac{ct}{\lambda_{\rm ij}} \int \dot{\epsilon}_{\rm ij}(x,y,z) dx dy\right] d\lambda
\end{equation}
where the sum \{i,j\} runs over all possible transitions, and the integration 
is over the $x-y$ plane at location 
$z = ct (\lambda_{ij} - \lambda)/\lambda_{ij}$.
The spectral flux observed at earth is simply 
$F_\lambda = L_\lambda/(4 \pi D^2)$, where $D$ 
is the distance to the source.

\subsection{Physical Processes Neglected}
\label{sssec:dust}

Dust formation might be possible in SNe~Ia at late times, but not necessarily in significant amounts \citep{nozawa2011}. Further, observations of the nearby SN011fe show no evidence for dust formation at 930 days past maximum \citep{kerzendorf2014}. We therefore neglect dust formation and plan on implementing it for future work. 

While implementing photoionization and stimulated radiative processes in our code, we neglect them for the work published in this paper under the assumption that there is negligible continuum radiation field. 

Charge transfer (CT) is expected to occur in SNe between neutral atoms and ions \citep{swartz1994}. While CT might affect ionization fractions of SNe~Ia at later times, we tested CT between ions of Fe and found that their effect on the nebular spectrum at 200-400 days is negligible. This is because the nebula is primarily ionized gas, and therefore the rate of ionization/recombination due to CT is sub-dominant. Since CT preferentially ionizes neutral atoms (as Coulomb repulsion suppresses ion-ion interactions), we expect that neutral atoms like OI and FeI should not contribute much to nebular emission. We plan to add a detailed CT treatment to the code in the near future in order to quantify this effect. 

\subsection{Code Verification}

We ran a number of tests to verify the code. We tested the NLTE level population solver in the limit that collisional processes dominate and the limit that the radiation field is the Planck function, both resulting in the expected LTE level populations. We also tested the ionization solution in the collisional-ionization equilibrium (CIE) regime and found good agreement with previous results \citep[e.g.][]{sutherland1993}. We further compared total CIE cooling functions for individual ions to published results from the Cloudy code \citep{gnat2012}.  Given disparities in the atomic data inputs, exact agreement of the cooling functions is not expected, but we found reasonably similar values and temperature dependences in the regimes of interest. The radiation transport calculation of synthetic spectra was verified by comparing single line profiles to analytic solutions.

\section{Modeling Type Ia nebular spectra} 
\label{sec:results}

To study the nebular spectra of SNe~Ia, we construct spherically symmetric  models in which the ejecta properties (e.g., total mass, energy, and abundances) are free parameters. We first describe the general properties of spectrum formation in a ``fiducial" model that resembles the ejecta structure expected for normal SNe~Ia. In \S\ref{sec:parameter_study} we carry out a parameter survey that demonstrates how the nebular spectra depend on explosion properties. 

\subsection{Ejecta Modeling}
\label{ssec:modeling}

We model the ejecta with a broken power-law density profile which is shallow in the core and steep in the outer layers \citep{chevalier1989,kasen2010}.

\begin{equation}
\label{eqn:broken_powerlaw}
\rho(v) =\left\{
	\begin{array}{ll}
     \rm \rho_0 \left(\frac{v}{v_t}\right)^{-\delta} & \rm ~v \leq v_t \\  
     \rm \rho_0 \left(\frac{v}{v_t}\right)^{-n} & \rm ~v > v_t \\
	\end{array} 
\right.
\end{equation}
where \sub{\rho}{0} can be interpreted as the central density of a perfectly flat core profile ($\rm \delta = 0$) and \sub{v}{t} is the transition velocity marking the interface of the two regions. Integration gives (assuming $\rm \delta < 3$ and $\rm n > 3$)
\begin{equation}
\label{eqn:rho0}
\rm \rho_0 = \frac{M_{ej}}{4 \pi (v_t t_{ex})^3} \left[\frac{1}{3 - \delta} + \frac{1}{n - 3}\right]^{-1}
\end{equation}

\begin{equation}
\label{eqn:E_K}
\rm E_K = \frac{1}{2}M_{ej}v_t^2 \left[\frac{1}{5 - \delta} + \frac{1}{n - 5}\right]\left[\frac{1}{3 - \delta} + \frac{1}{n - 3}\right]^{-1}
\end{equation}
where \sub{M}{ej} is the total ejecta mass, $E_{\rm k}$ is the ejecta kinetic energy, and \sub{t}{ex} is the time since explosion.  The radial density profile is thus completely set by the choice of $M_{\rm ej},~E_{\rm k}$, and the exponents
$\delta$, $n$. In our calculations, we cut off the model at a radius that encompasses 99\% of the total ejecta mass.

We find  that the values $\rm \delta = 0,~n=10$ give reasonable fits to the nebular spectra of SNe~Ia, and so use these values for our fiducial model. In this case, the characteristic velocity and density scales are 

\begin{equation}
\label{eqn:v_t}
\rm v_t =  10,943 ~E_{51}^{1/2} ~M_{1}^{-1/2}~km~s^{-1}
\end{equation}

\begin{equation}
\label{eqn:rho0_delta}
\rm \rho_0 = 4.90 \times 10^{-17} ~E_{51}^{-3/2} ~M_{1}^{5/2} ~t_{200}^{-3} ~g~cm^{-3}
\end{equation}
where $M_{1} = M_{\rm ej}/M_\odot$, $E_{51}$ is the kinetic energy in units of $10^{51}$ erg, and \sub{t}{200}\ is the time since explosion scaled to 200 days.  We explore using different power law exponents, as well as an exponential density profile, in \S\ref{sec:density_profiles}.

The compositional structure of the ejecta models is assumed to be stratified into three distinct zones \citep{woosley2007}. The center of the ejecta is assumed to be consist of stable iron group elements (IGEs) of mass $M_{\rm IGE}$. The stable IGEs are assumed to be composed of a ratio, \sub{R}{stb}, of $^{54}$Fe to $^{58}$Ni. Surrounding the stable IGE region is a zone consisting (initially) mostly of \fsix{Ni} of mass \sub{M}{56Ni}. We include a small amount of stable IGEs in this region with mass abundance \sub{X}{stb} and the same isotopic ratio \sub{R}{stb}. Above the \fsix{Ni} is an outer layer of intermediate mass elements (IMEs) of mass \sub{M}{IME}. We compose the IME layer of 70\% $^{28}$Si, 29\% $^{32}$S, and 1\% $^{40}$Ca, roughly consistent with the nucleosynthetic results in the SNIa explosion models of \citet{plewa2007} and \citet{seitenzahl2013}. We study the presence of unburned C/O mixed into the nickel zone and IME layer in \S\ref{ssec:M_CO}. The parameters describing the masses of the elements are constrained to add to the total ejecta mass. 

The total radioactive energy deposition rate (and hence bolometric luminosity) of a model depends not only on \sub{M}{56Ni} but also on the efficiency of the trapping of radioactive decay products. Since the ejecta are largely transparent to gamma rays at nebular phases, the gamma-ray trapping fraction is $f_{\rm \gamma,c} = \rm 1 - e^{-\tau_\gamma} \approx  \tau_\gamma$, where $\tau_\gamma$ is the mean optical depth to gamma-rays. Taking a typical gamma-ray opacity $ \rm \kappa_\gamma \sim 0.03~cm^2~g^{-1}$ \citep{swartz1995} and integrating the  radial optical depth from the center ($r=0$) gives an estimate
\begin{equation}
\label{eqn:f_gamma_final}
\rm f_{\gamma,c} \approx 0.025 ~E_{51}^{-1} ~M_{1}^2 ~t_{200}^{-2}
\end{equation}
Eq.~\ref{eqn:f_gamma_final} presumably overestimates the gamma-ray trapping fraction (since $\tau_\gamma$ is evaluated at $r=0$) but the scaling of $f_{\gamma,c}$ with \sub{M}{ej},  \sub{E}{K}, and $t$ will be useful for interpreting how radioactive deposition rate depends on physical parameters.

\subsection{Fiducial Model}
\label{ssec:fiducial}

To describe the basic features of nebular spectrum formation, we present first a fiducial model with parameters (given in Table \ref{table:fiducial}) typical of standard SN~Ia explosion models, i.e., $M_{\rm ej}$ near the Chandrasekhar mass and $M_{\rm 56Ni} = 0.6~M_\odot$. The fiducial model has a transition velocity of $\rm 10,131~km~s^{-1}$ while the interface between the nickel zone and IME layer is near $\rm 8,800~km~s^{-1}$. 

\begin{table}[ht]
\label{table:fiducial}
\centering
\begin{tabular}{|c|c|}
\hline
\sub{t}{ex}\ (days) & 200 \T \\
\hline
\sub{M}{ej}\ (\Msun) &  1.40 \T \\
\hline
\sub{E}{K}\ ($10^{51}$ erg) & 1.2 \T \\
\hline
\sub{M}{56Ni}\ (\Msun)  & 0.6 \T \\
\hline
\sub{M}{stb}\ (\Msun) & 0.0 \T \\
\hline
\sub{X}{stb} & 0.05 \T \\ 
\hline
\sub{R}{stb} & 1.0 \T \\
\hline
\sub{M}{IME} (\Msun) & 0.75 \T \\
\hline
\sub{M}{CO} (\Msun) & 0.0 \T \\
\hline
\end{tabular}
\caption{Fiducial model parameters. \sub{M}{stb}\ refers to the mass of the stable IGE region in the core of the explosion; \sub{X}{stb}\ is the mass fraction of stable IGE material mixed into the \fsix{Ni} region; \sub{R}{stb}\ is the ratio of $^{54}$Fe/$^{58}$Ni in both core stable IGE and \fsix{Ni} regions; \sub{M}{IME} is the total mass of the IME layer; \sub{M}{CO}\ is the mass of C/O mixed throughout the ejecta.}
\end{table}

\begin{figure}[htbp]
\includegraphics[width=0.5\textwidth]{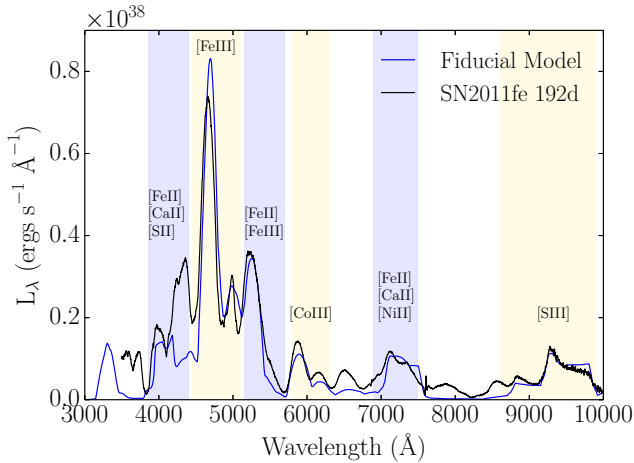}
\label{fig:fid_spec_2011fe}
\caption{Synthetic spectrum of the fiducial SN~Ia model (at 200 days after explosion) compared to SN2011fe at 192 days \citep{mazzali2015}.}
\end{figure}

Figure \ref{fig:fid_spec_2011fe} shows our calculation of the synthetic nebular spectrum of the fiducial model at 200 days after explosion. We compare to the observed spectrum of the well-studied nearby Type~Ia SN2011fe \citep{nugent2011, shappee2013, mazzali2015}. The model spectrum reproduces most of the prominent features.

\begin{figure}[htbp]
\includegraphics[width=0.5\textwidth]{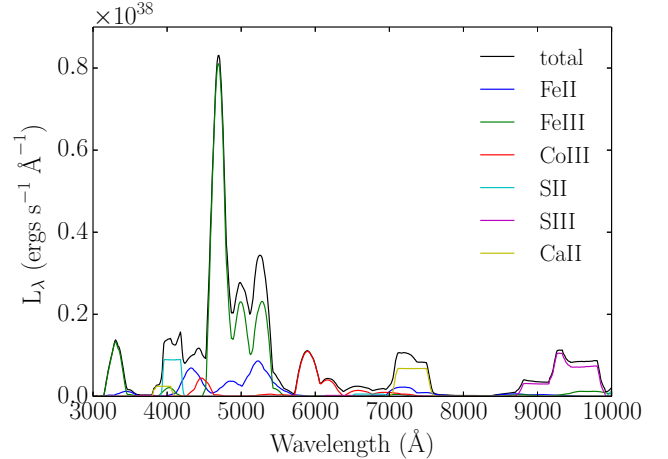}
\label{fig:fid_decomp}
\caption{Fiducial model spectrum decomposed into the emission from individual ions. Forbidden transitions of FeIII dominate the spectrum, along with some FeII, CoIII, CaII, SII, and SIII features.}
\end{figure}

Figure \ref{fig:fid_decomp} shows a breakdown of the contribution from various ions to the fiducial model spectrum. The strongest features are due to forbidden transitions of FeII and FeIII which are collisionally-excited in the nickel zone. Emission due to IME lines is also visible at redder wavelengths. Most of the important individual line transitions are listed in Table \ref{table:transitions}.

The spectrum in the nebular phase forms primarily from the  collisional excitation of ions by thermal electrons, followed by spontaneous de-excitation via the emission of a line photon. Given the relatively low ejecta temperatures, electrons can only excite low-lying atomic levels, among which radiative transitions are typically electron dipole forbidden (see Table~\ref{table:transitions}). Nevertheless, the rate of collisional de-excitations is generally so small in the low-density nebula that essentially every collisional excitation eventually leads to radiative emission through a forbidden line.

\begin{figure}[htbp]
\includegraphics[width=0.5\textwidth]{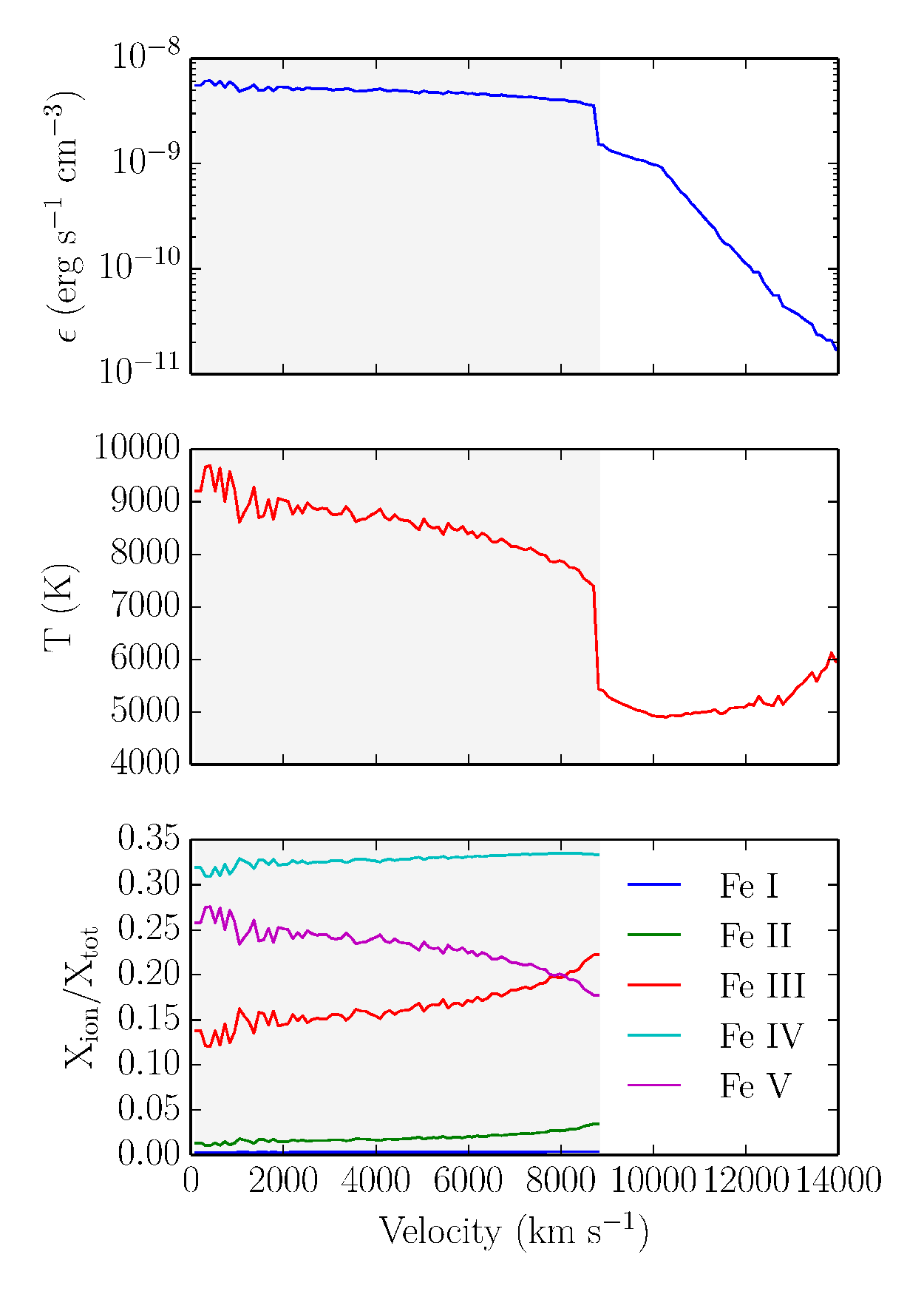}
\label{fig:fid_radial}
\caption{Fiducial model calculated properties. The top panel shows energy deposition rate from \fsix{Co} decay, including gamma-ray and positron channels. The middle panel shows calculated temperature based on the balance of \fsix{Co} heating and line cooling. The bottom panel shows the ionization fractions of FeI-V. The nickel zone, which contains only IGEs, is shaded in gray. The discontinuity at $\sim$8800 $\rm km~s^{-1}$ is due to the interface between the nickel zone and IME layer.}
\end{figure}

The temperature of the ejecta is determined by the balance of radioactive heating and cooling by line emission. Figure \ref{fig:fid_radial} shows the radioactive heating rate and temperature for the fiducial model. The interior ejecta density profile is flat in this model, and so the ejecta temperature is nearly constant at $T \approx 9000$~K in the inner layers. Above the radioactive nickel zone, the temperature drops due to the declining heating rate, but increases again in the very low-density outermost layers due to the inability of the ejecta to cool efficiently. 

The important emission lines appearing in the nebular spectra depend on the composition and ionization state of the ejecta.  
In SNe~Ia, ionization is primarily caused by the non-thermal electrons produced by radioactive decay (the collisional ionization rate from thermal electrons is sub-dominant), which is balanced by radiative recombination.  Figure \ref{fig:fid_radial} shows the radial dependence of the ionization fractions of iron for the fiducial model. Though FeIV and FeV are the most abundant ions, they lack low lying levels that are able to be excited by the thermal electrons. Thus the most prominent lines are due to FeIII and FeII. In general, the degree of ionization increases as the density declines, reflecting the reduced rate of radiative recombination.

As a more comprehensive description of SN~Ia nebular
spectra, we discuss the features appearing in each key wavelength region seen in Figure~\ref{fig:fid_spec_2011fe}:

\smallskip \noindent \textbf{3500-4500 \AA} This region contains emission from [SII] and [FeII] transitions. Similar to other studies \citep[e.g.][]{mazzali2015}, our model fails to reproduce all of the observed features, which could be a result of ions missing in the model, uncertainties or incompleteness of the atomic line data, or the neglect of optical depth effects that may produce a pseudo-continuum at the bluest wavelengths.

\smallskip \noindent \textbf{4500-5500 \AA} This region is dominated by emission from [FeIII], which produces prominent features at 4658 \AA\ and 5270 \AA. The latter feature also includes a significant contribution from [FeII] 5159 \AA; therefore, the ratio of these two lines is a diagnostic of the ionization state of the gas. The small emission line appearing between the two strong lines is due to [FeIII] 5011 \AA, and can be washed out if the ejecta velocities are too high.

\smallskip \noindent \textbf{5500-7000 \AA} This region contains a prominent [CoIII] feature near 5888 \AA\ with two smaller [CoIII] features immediately to the red. The line strength depends on the abundance and ionization state of Cobalt, and typically declines with time as \fsix{Co} decays to \fsix{Fe}. The feature near 6500 \AA\ is not well fit by our model (or by previous models e.g., \citet{mazzali2015}). Because stripped hydrogen from a non-degenerate companion should have velocities $\rm \lesssim ~ 1000~km~s^{-1}$ \citep[e.g.][]{marietta2000}, this emission is unlikely to be due to H$\alpha$. The identification of the feature and the reason for the poor fit are thus unclear.

\smallskip \noindent \textbf{7000-7700 \AA} The feature in this region is a blend of multiple [FeII] lines and a broad [CaII] line. In addition, [NiII] emission can contribute if sufficient stable $^{58}$Ni is present in the gas, but for our fiducial model [NiII] does not dominate this feature. Notably, the fiducial model only contains 0.0075\Msun\ of calcium and yet [CaII] emission from the IME layer dominates the emission in this region. 

\smallskip \noindent \textbf{8500-11000 \AA} This region is made up primarily of [SII] and [SIII] emitted by the 0.22\Msun\ of sulfur in the IME layer above the nickel zone. The broad, flat-topped profiles of IMEs are due to their large velocities and absence of IMEs in the core.

\begin{table}[htb]
\label{table:transitions}

\begin{center}
\caption{Line Identifications for Optical SNe~Ia Nebular Spectra}
\resizebox{\columnwidth}{!}{
\begin{tabular}{|c|c|c|c|c|c|c|}
\hline
\textbf{Ion} & \textbf{$ \rm \lambda_{ij}$ ( \AA)} & \textbf{$\rm A_{ij}$ (s$^{-1}$)} & \textbf{Conf. ($\rm ^{2S+1}L^\pi_J$)} & $\rm E_i $ (eV) & $\rm E_j$ (eV) \T \\
\hline

SII & 4069 & 0.2250 & $ \rm S^o_{3/2}-^2P^o_{3/2}$ & 0.00 & 3.05 \T \\   
FeIII & 4658 & 0.4500 & $ \rm ^5D^e_4-^3_2F2^e_4 $ & 0.00 & 2.66 \T  \\ 
OIII & 5007 & 0.0196 & $ \rm ^3P^e_2-^1D^e_2 $ & 0.04 & 2.51 \T  \\ 
FeIII & 5011 & 0.5400 & $ \rm ^5D^e_2-^3P2^e_1 $ & 0.09 & 2.57 \T \\
FeII & 5159 & 0.5805 & $ \rm a^4F^e_{9/2}-a^4H^e_{13/2} $ & 0.23 & 2.63 \T \\
FeIII & 5270 & 0.4200 & $ \rm ^5D^e_3-^3P2^e_2$ & 0.05 & 2.41 \T \\

CoIII & 5888 & 0.4001 & $ \rm a^4F^e_{9/2}-a^2G^e_{9/2}$ & 0.00 & 2.10 \T \\
CoIII & 6128 & 0.1100 & $ \rm a^4F^e_{5/2}-a^2G^e_{7/2}$ & 0.18 & 2.20 \T \\
CoIII & 6195 & 0.1200 & $ \rm a^4F^e_{7/2}-a^2G^e_{9/2}$ & 0.10 & 2.10 \T \\

FeII & 7155 & 0.1495 & $\rm a^4F^e_{9/2}-a^2G^e_{9/2}$ & 0.23 & 1.96 \T \\
CaII & 7291 & 0.803 & $\rm ^2S^e_{1/2}-^2D^e_{5/2}$ & 0.00 & 1.70 \T \\
NiII & 7378 & 0.1955 & $\rm ^2D^e_{5/2}-^2F^e_{7/2}$ & 0.00 & 1.16 \T \\
FeII & 7388 & 0.0435 & $\rm ^4F^e_{5/2}-a^2G^e_{7/2}$ & 0.35 & 2.03 \T \\
FeII & 7453 & 0.0485 & $\rm a^4F^e_{7/2}-a^2G^e_{9/2}$ & 0.30 & 1.96 \T \\
FeII & 7638 & 0.0070 & $\rm a^6D^e_{7/2}-a^4P^e_{5/2}$ & 0.05 & 1.67 \T \\

FeII & 8617 & 0.0334 & $\rm a^4F^e_{9/2}-a^4P^e_{5/2}$ & 0.23 & 1.67 \T \\   
FeIII & 8729 & 0.0495 & $\rm ^3P2^e_2-^3D^e_3$ & 2.41 & 3.83 \T \\     
SIII & 9069 & 0.0221 & $\rm ^3P^e_1-^1D^e_2$ & 0.04 & 1.40 \T \\
SIII & 9531 & 0.0576 & $\rm ^3P^e_2-^1D^e_2 $ & 0.10 & 1.40 \T \\

SII & 10336 & 0.1630 & $\rm ^2D^o_{3/2}-^2P^o_{1/2}$ & 1.84 & 3.04 \T \\
SII & 10370 & 0.0779 & $\rm ^2D^o_{5/2}-^2P^o_{1/2}$ & 1.85 & 3.04 \T \\

\hline

\end{tabular}
}
\end{center}
\end{table}

\section{Systematic Parameter Study}
\label{sec:parameter_study}

We present the following systematic study probing the sensitivity of synthetic nebular spectra to model parameters and atomic data inputs. In particular, we calculate how the spectra depend on the total ejecta mass, kinetic energy, \fsix{Ni} mass, and time since explosion, as well as density and compositional structure. A full list of models and their derived properties is shown in Table \ref{table:models}. 

\begin{table*}[!htbp]
\centering
\label{table:models}
\begin{center}
\begin{tabular}{|c|c|c|c|c|c|c|c|c|c|}

\hline
\textbf{Model} & \textbf{\sub{v}{t}} & \textbf{\sub{\rho}{c}} & \textbf{\sub{M}{IME}} & \textbf{\sub{M}{stb,tot}} & \textbf{\sub{L}{bol}} & \textbf{L$_{4665}$/L$_{5272}$} \T \\

& (km/s) & ($10^{-17}$ g/cm$^{3}$) & (\Msun)& Fe, Ni (\Msun) & (10$\rm ^{40}~erg/s$) &  \\
\hline

fiducial & 10,131 & 8.65 & 0.75 & 0.016, 0.016 & 7.96 & 2.30 \T \\
\hline
$t\rm _{ex} = 250~days$ & 10,131 & 4.43 & 0.75 & 0.016, 0.016 & 3.93 & 2.45 \T \\
$t\rm _{ex} = 200~days$ & 10,131 & 2.56 & 0.75 & 0.016, 0.016 & 2.11 & 2.53 \T \\
$t\rm _{ex} = 350~days$ & 10,131 & 1.61 & 0.75 & 0.016, 0.016 & 1.19 & 2.54 \T \\
$t\rm _{ex} = 400~days$ & 10,131 & 1.08 & 0.75 & 0.016, 0.016 & 0.70 & 2.48 \T \\
\hline
$M\rm _{ej} = 1.0~M_\odot$ & 12,000 & 3.73 & 0.36 & 0.016, 0.016 & 5.27 & 2.75 \T \\
$M\rm _{ej} = 1.1~M_\odot$ & 11,442 & 4.73 & 0.46 & 0.016, 0.016 & 5.82 & 2.65 \T \\
$M\rm _{ej} = 1.2~M_\odot$ & 10,954 & 5.88 & 0.54 & 0.016, 0.016 & 6.58 & 2.53 \T \\
$M\rm _{ej} = 1.3~M_\odot$ & 10,525 & 7.19 & 0.65 & 0.016, 0.016 & 7.18 & 2.41 \T \\
$M\rm _{ej} = 2.0~M_\odot$ & 8,485 & 21.10 & 1.33 & 0.016, 0.016 & 13.91 & 1.86 \T \\
\hline
$E\rm _K = 1.0B$ & 9,258 & 11.37 & 0.75 & 0.016, 0.016 & 8.90 & 2.15 \T \\
$E\rm _K = 1.1B$ & 9,710 & 9.86 & 0.75 & 0.016, 0.016 & 8.33 & 2.23 \T \\
$E\rm _K = 1.3B$ & 10,556 & 7.67 & 0.75 & 0.016, 0.016 & 7.57 & 2.37 \T \\
$E\rm _K = 1.4B$ & 10,954 & 6.86 & 0.75 & 0.016, 0.016 & 7.22 & 2.44 \T \\
$E\rm _K = 1.5B$ & 11,339 & 6.19 & 0.75 & 0.016, 0.016 & 6.96 & 2.50 \T \\
$E\rm _K = 2.0B$ & 13,093 & 4.02 & 0.75 & 0.016, 0.016 & 5.79 & 2.73 \T \\
\hline
$M\rm _{56Ni} = 0.4~M_\odot$ & 10,131 & 8.65 & 0.96 & 0.011, 0.011 & 5.49 & 2.25 \T \\
$M\rm _{56Ni} = 0.5~M_\odot$ & 10,131 & 8.65 & 0.85 & 0.013, 0.013 & 6.77 & 2.28 \T \\
$M\rm _{56Ni} = 0.7~M_\odot$ & 10,131 & 8.65 & 0.64 & 0.019, 0.019 & 9.22 & 2.32 \T \\
$M\rm _{56Ni} = 0.8~M_\odot$ & 10,131 & 8.65 & 0.55 & 0.021, 0.021 & 10.21 & 2.33 \T \\
\hline
$M\rm _{stb} = 0.05~M_\odot$ & 10,131 & 8.65 & 0.71 & 0.040, 0.040 & 7.75 & 2.28 \T \\
$M\rm _{stb} = 0.10~M_\odot$ & 10,131 & 8.65 & 0.64 & 0.066, 0.066 & 7.81 & 2.25 \T \\
$M\rm _{stb} = 0.15~M_\odot$ & 10,131 & 8.65 & 0.59 & 0.093, 0.093 & 7.64 & 2.21 \T \\
$M\rm _{stb} = 0.20~M_\odot$ & 10,131 & 8.65 & 0.54 & 0.116, 0.116 & 7.55 & 2.17 \T \\
\hline
$M\rm _{ej} = 1.0~scaled\footnote{In these models, we keep the following ratios fixed: $\rm M_{56Ni}/M_{ej} = 1/2$ and $\rm E_{51}/M_{ej} = 1.2/1.4$. See \S\ref{ssec:M_ej} for further details.}$ & 10,240 & 6.15 & 0.46 & 0.013,0.013 & 5.35 &  2.49 \T \\
$M\rm _{ej} = 1.4~scaled$ & 10,240 & 8.65 & 0.64 & 0.019,0.019 & 9.21 & 2.32 \T \\
$M\rm _{ej} = 2.0~scaled$ & 10,240 & 12.40 & 0.91 & 0.027,0.027 & 16.88 & 2.17 \T \\
\hline
$M\rm _{CO} = 0.05~M_\odot$ & 10,131 & 8.65 & 0.70 & 0.016, 0.016 & 7.85 & 2.25 \T \\
$M\rm _{CO} = 0.10~M_\odot$ & 10,131 & 8.65 & 0.64 & 0.017, 0.017 & 8.04 & 2.20 \T \\
$M\rm _{CO} = 0.15~M_\odot$ & 10,131 & 8.65 & 0.59 & 0.018, 0.018 & 7.90 & 2.15 \T \\
$M\rm _{CO} = 0.20~M_\odot$ & 10,131 & 8.65 & 0.55 & 0.019, 0.019 & 7.82 & 2.09 \T \\
$M\rm _{CO} = 0.40~M_\odot$ & 10,131 & 8.65 & 0.33 & 0.023, 0.023 & 7.76 & 1.86 \T \\
\hline
$\delta = 0.5$, n = 10\footnote{These refer to the power-law exponents in Equation \ref{eqn:broken_powerlaw}. A higher value of $\delta$ produces a steeper density profile.} 
& 10,528 & 66.89 & 0.75 & 0.016, 0.016 & 8.21 & 2.17 \T \\
$\delta = 1.0$, n = 10 & 11,098 & 486.5  & 0.75 & 0.016, 0.016 & 8.95 & 1.88 \T \\
exponential\footnote{This refers to the exponential density profile explained in \S\ref{sec:density_profiles}.} & N/A &  106.6 & 0.75 & 0.016, 0.016 & 12.56 & 1.63 \T \\
\hline
all $Q\rm _{k} \times 2$\footnote{$Q\rm_{k}$ is the collisional (electron impact) ionization cross section for the ground state of ion $k$.} & 10,131 & 8.65 & 0.75 & 0.016, 0.016 & 7.96 & 2.49 \T \\
all $Q\rm _{k} \times 0.5$ & 10,131 & 8.65 & 0.75 & 0.016, 0.016 & 7.96 & 2.02 \T \\
$Q\rm _{FeII} \times 2$, $Q\rm _{FeIII} \times 0.5$ & 10,131 & 8.65 & 0.75 & 0.016, 0.016 & 7.96 & 2.74 \T \\
$Q\rm _{FeII} \times 0.5$, $Q\rm _{FeIII} \times 2$ & 10,131 & 8.65 & 0.75 & 0.016, 0.016 & 7.96 & 1.72 \T \\
all $\sigma \rm _{ij} \times 2$\footnote{$\sigma\rm _{ij}$ is the thermal collisional (electron-impact) excitation rate for a transition $ij$.} & 10,131 & 8.65 & 0.75 & 0.016, 0.016 & 7.96 & 2.10 \T \\
all $\sigma \rm _{ij} \times 0.5$ & 10,131 & 8.65 & 0.75 & 0.016, 0.016 & 7.96 & 2.52 \T \\
\hline
$\alpha\rm ~from~NORAD$ & 10,131 & 8.65 & 0.75 & 0.016, 0.016 & 7.96 & 1.66 \T \\
\hline
\end{tabular}
\caption{List of models included in physical parameter study. The parameter varied from the fiducial value in each model is shown in the model description, and some derived values are shown. \sub{v}{t}\ is the transition velocity of the density profile determined by power-law exponents $\delta$ and n. $\rm \rho_c$ is the central density. The column for \sub{M}{stb,tot}, which is the sum of neutron-rich core material and stable isotopes mixed into the nickel zone, reports $^{54}$Fe and $^{58}$Ni masses separately. \sub{L}{bol}\ is the total luminosity over all wavelengths. $\rm L_{4665}/L_{5272}$ is the ratio of luminosities at 4665 \AA\ and 5272 \AA, the wavelengths at which the prominent [FeIII] and [FeII]/[FeIII] features peak, respectively.}
\end{center}
\end{table*}

\subsection{Time since explosion}
\label{ssec:t}

\begin{figure*}[htbp]
\includegraphics[width=0.99\textwidth]{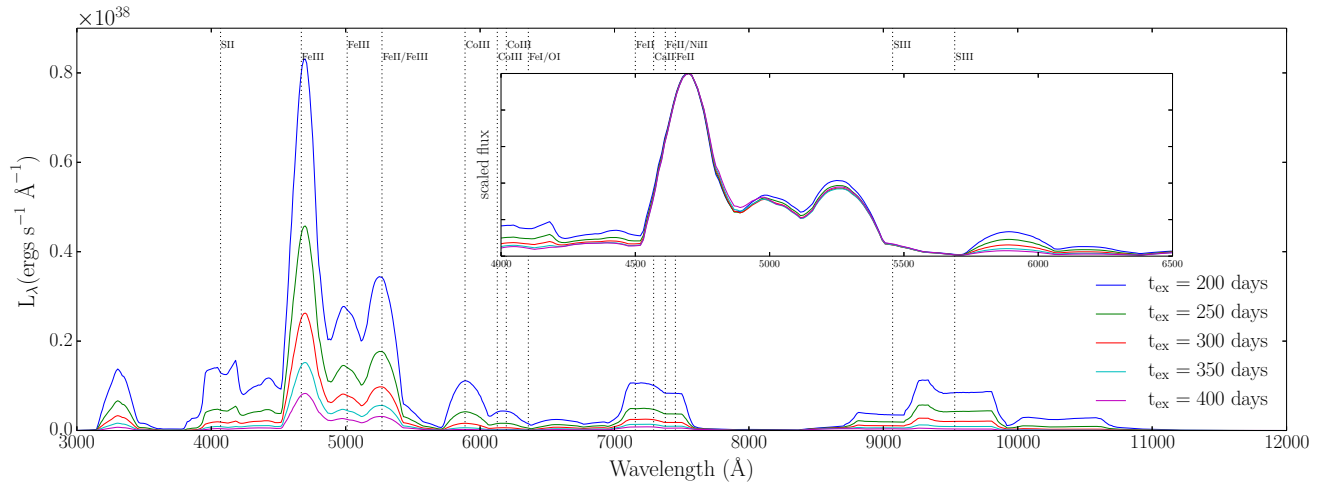}
\label{fig:t_spec_scaled}
\caption{Synthetic spectra of the models with varying time since explosion. The inset shows all fluxes scaled to the peak of the 4658 \AA\ feature. Vertical dotted lines identify the source of some prominent transitions. Over time, the bolometric luminosity declines due to decreased energy deposition, and the [CoIII] 5888~\AA\ feature becomes weaker due to the decay of radioactive \fsix{Co}. }
\end{figure*}

Figure~\ref{fig:t_spec_scaled} shows the nebular spectrum of the fiducial model at times between 200 and 400 days after explosion. As the supernova evolves in time, ejecta densities decline and radioactive isotopes decay. The bolometric luminosity drops with time due to the declining radioactive heating and a decreasing gamma-ray trapping fraction. The relative strength of features in the spectrum, however, remains fairly constant, as seen in the figure inset. The ratio of the prominent Fe lines does not evolve significantly, indicating that the FeIII/FeII ionization ratio remains fairly constant over time. 

The biggest change in the spectral features over time is the strength of the [CoIII] emission at 5888 \AA, which decreases as \fsix{Co} decays to \fsix{Fe}. In addition, the IME emission of [SII] 4000 \AA\ declines at later times as a result of a declining sulfur ionization fraction in the IME layer.

\subsection{Explosion kinetic energy}
\label{ssec:E_K}
The amount of kinetic energy (\sub{E}{K}) imparted to the ejecta by a supernova explosion depends on the nucleosynthetic yields and initial binding energy \citep{woosley2007}. The typical energy of SNe~Ia is around a Bethe (1B = $\rm 10^{51}~erg$). Various theoretical models have predicted kinetic energies in the range $0.87 - 1.6$~B \citep{gamezo2005, golombek2005, plewa2007, ropke2007a,  jordan2008, ropke2007b, bravo2009b, jordan2012}.  

\begin{figure*}[htbp]
\includegraphics[width=0.99\textwidth]{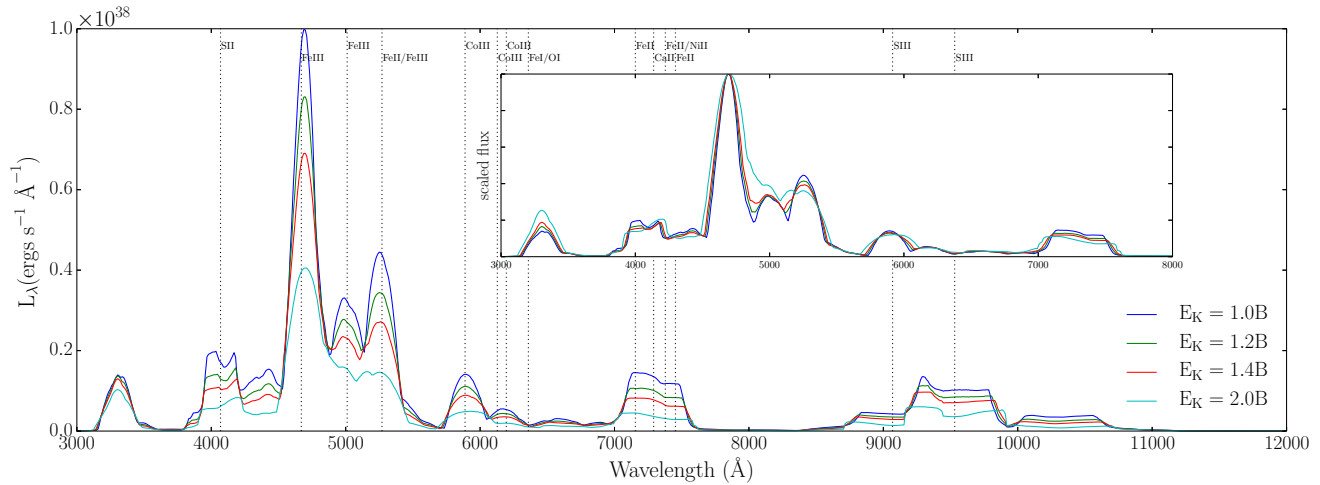}
\label{fig:E_K_specs}
\caption{Same as Figure \ref{fig:t_spec_scaled} but for the \sub{E}{K}\ parameter study. Increasing \sub{E}{K} produces slightly wider line profiles, eventually washing out features like the [FeIII] emission at 5011 \AA. Bolometric luminosity drops as higher \sub{E}{K} models have decreased energy deposition. }
\end{figure*}

Figure \ref{fig:E_K_specs} shows synthetic spectra of the fiducial model with \sub{E}{K}\ varied between $1 - 2$~B. The lower \sub{E}{K}\ models are more efficient at trapping radioactive energy (due to the higher ejecta density, Eq.~\ref{eqn:rho0_delta}) and so have  higher bolometric luminosities. 

Changing \sub{E}{K} has a modest effect on the shape of the spectral features. An increase of \sub{E}{K} from 1~B to 1.4~B increases the velocity scale by only 18\%, which results in a subtle increase in the line widths (as these widths are also set in part by line blending). A larger change of \sub{E}{K} by a factor of 2 (from 1~B to 2~B) does have noticeable effects, causing the FeIII/FeII complex to be so blended that the small central emission near 5000 \AA\ becomes indistinguishable. Observations of this small feature may therefore be a useful diagnostic of the velocity of the nickel zone in SNe~Ia.

\subsection{Total ejecta mass}
\label{ssec:M_ej}

Theoretical models of SNe~Ia predict an ejected mass in the range $0.8 - 2.0$~\Msun, depending on the progenitor scenario. Approximate light curve modeling studies have suggested that observed SNe~Ia could span this entire range \citep{stritzinger2006a, scalzo2010, scalzo2014a, scalzo2014b}. 

\begin{figure*}[htbp]
\centering
\includegraphics[width=0.99\textwidth]{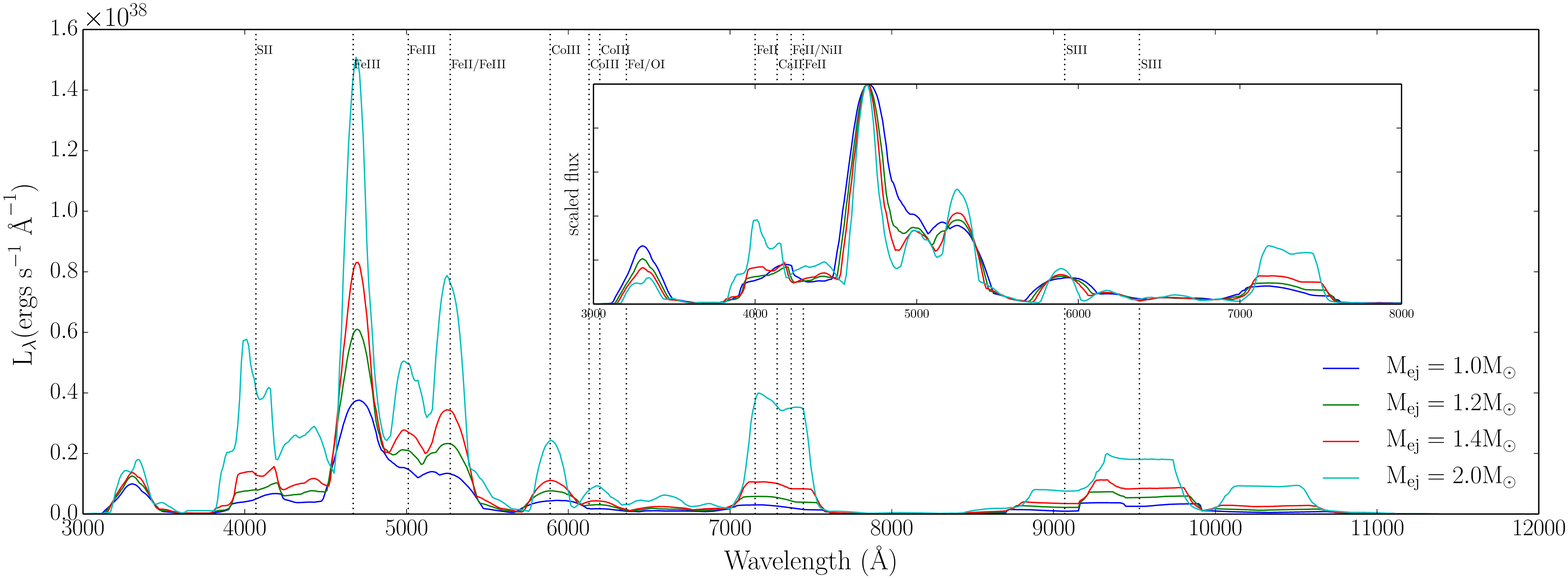}
\label{fig:M_ej_spec_scaled}
\caption{Same as Figure \ref{fig:t_spec_scaled} but for the \sub{M}{ej}\ parameter study (kinetic energy and \fsix{Ni} mass are held fixed). Increasing the total ejecta mass results in higher IME emission around 4000 \AA, 7300 \AA, and 9500 \AA\ given that those models have higher IME masses by construction. The effect of reducing \sub{M}{ej} is largely degenerate with increasing the kinetic energy (Figure \ref{fig:E_K_specs}) with a low ejecta mass producing a highly-blended feature around 5007 \AA. }
\end{figure*}

Figure \ref{fig:M_ej_spec_scaled} shows synthetic spectra of the fiducial model with \sub{M}{ej}\ varied between $1 - 2$~\Msun. For a fixed kinetic energy, a higher \sub{M}{ej}\ results in higher ejecta densities and lower velocities. As a result, higher \sub{M}{ej}\ models have greater gamma-ray trapping, a brighter bolometric luminosity, and less Doppler broadened spectral features.  This effect of \sub{M}{ej}\ is therefore somewhat degenerate with that of kinetic energy. 

The relative features in the synthetic spectra of
Figure \ref{fig:M_ej_spec_scaled} show only subtle variations with \sub{M}{ej}. For our super-\sub{M}{Ch} case with $M_{\rm ej} = 2.0~M_\odot$, the IME features become visibly stronger due to the higher total IME mass. The FeIII/FeII line ratio also decreases due to the increased rate of recombination at higher densities.  

While we have studied the effect of varying ejecta mass alone in Figure \ref{fig:M_ej_spec_scaled}, this parameter is likely correlated with kinetic energy and \fsix{Ni} mass \citep{woosley2007}; for example, a super-\sub{M}{Ch} explosion is likely to produce more \fsix{Ni} and higher kinetic energy. We therefore ran three additional models in which we fixed the ratios $\rm M_{56Ni}/M_{ej} = 1/2$ and $\rm E_{51}/M_{ej} = 1.2/1.4$. Based on the above discussion of gamma-ray trapping (\S\ref{ssec:modeling}), we expect luminosity to approximately scale as $\rm L_{bol} \sim M_{ej}^2$ if these ratios are held constant. 

\begin{figure*}[htbp]
\includegraphics[width=0.99\textwidth]{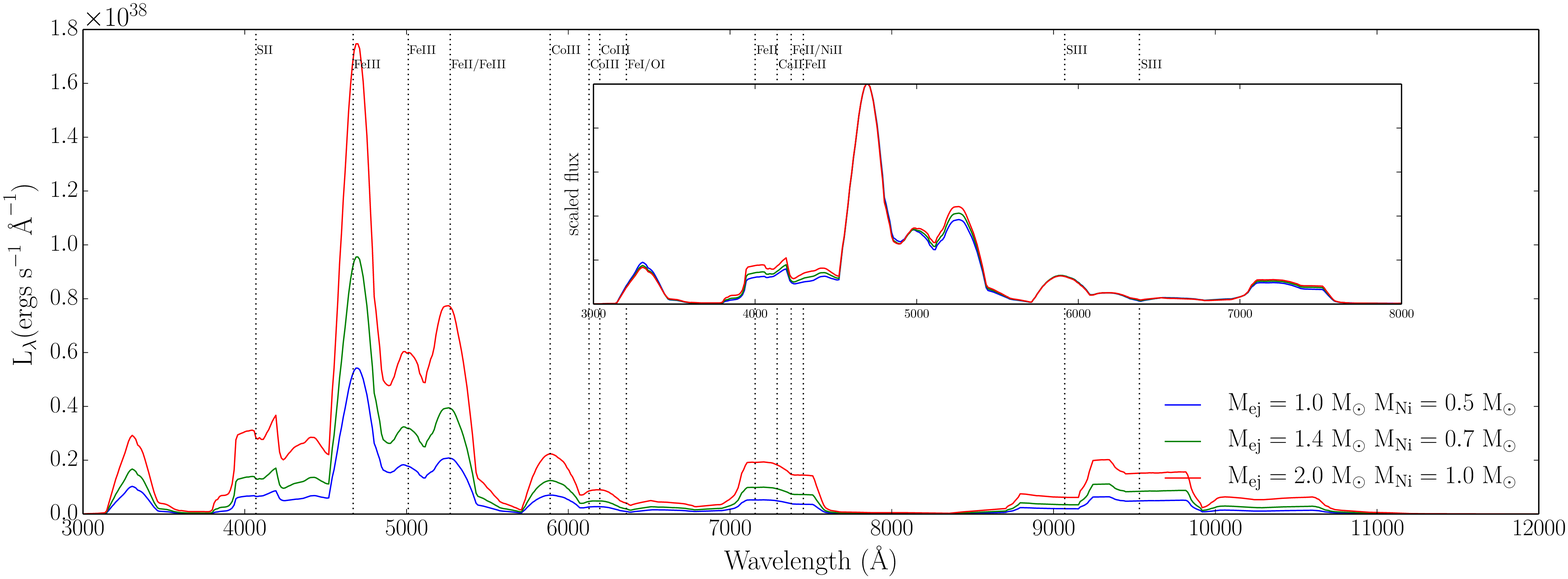}
\label{fig:M_scl_spec}
\caption{Same as Figure \ref{fig:t_spec_scaled} but for the scaled parameter study which varies $M_{\rm ej}$ while keeping a fixed ratio of $\rm M_{56Ni}/M_{\rm ej} = 1/2$ and $\rm E_{51}/M_{\rm ej} = 1.2/1.4$. Unlike the models in which ejecta mass, kinetic energy, and \fsix{Ni} mass are varied individually, these models produce almost identical synthetic spectra in all respects except for the overall bolometric luminosity. }
\end{figure*}

Figure \ref{fig:M_scl_spec} shows synthetic spectra of these scaled model with total ejecta masses 1.0, 1.4, and 2.0~$M_\odot$. We find that the spectral features are remarkably unchanged despite a substantial variation in
the masses. The only major difference is the bolometric luminosity. This indicates that, in a generic sense, the nebular spectra of SNe~Ia are consistent with non-\sub{M}{Ch} models, provided \sub{E}{k} and \sub{M}{56Ni} scale accordingly.

\subsection{Radioactive nickel mass}
\label{ssec:M_56Ni}

A number of observational studies and theoretical models indicate that the \fsix{Ni} masses of normal SNe~Ia range from 0.3 to 1.2\Msun\ \citep{gamezo2005, mazzali2007a_hydro, ropke2007a, ropke2007b, plewa2007, jordan2008, bravo2009b, raskin2009, rosswog2009, seitenzahl2011, jordan2012, seitenzahl2013}, with a typical value near 0.6\Msun\ \citep{branch1995}.

\begin{figure*}[htbp]
\includegraphics[width=0.99\textwidth]{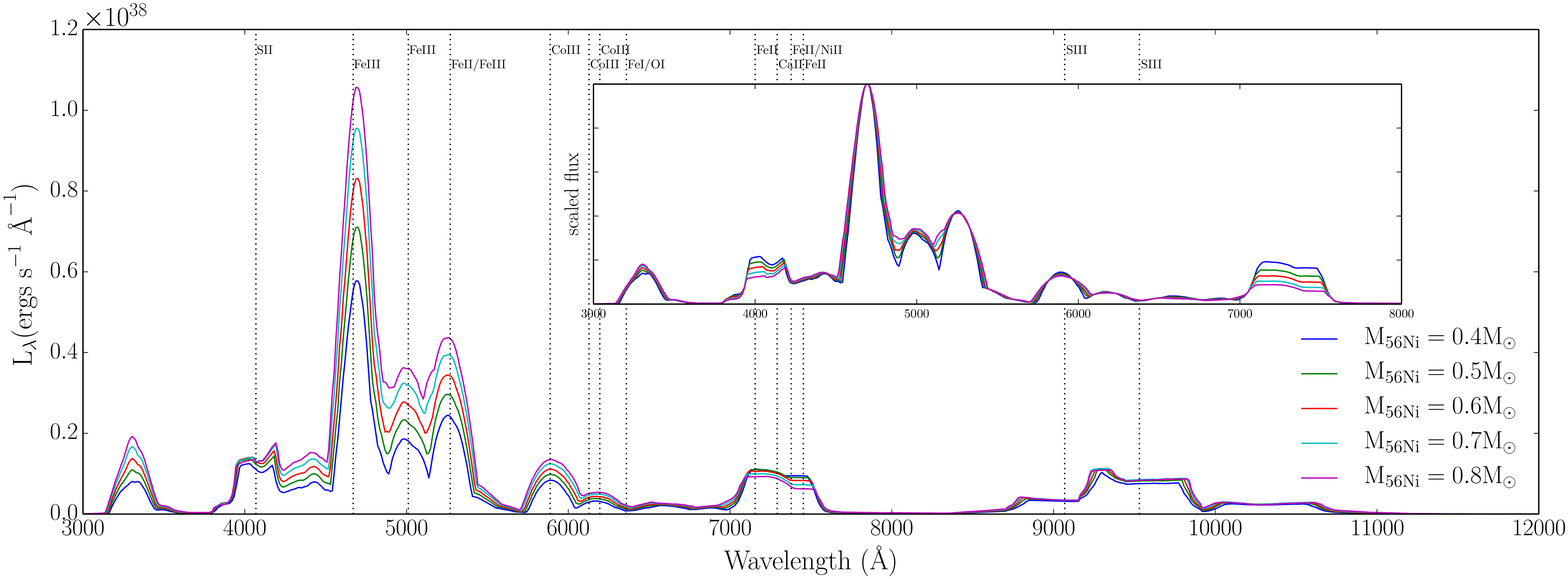}
\label{fig:M_56_spec_scaled}
\caption{Same as Figure \ref{fig:t_spec_scaled} but for the \sub{M}{56Ni}\ parameter study. Higher nickel mass models have larger nickel cores, and so produce slightly wider and more blended iron group features. The declining IME mass (a result of keeping \sub{M}{ej} fixed while increasing \sub{M}{56Ni}) results in declining IME emission.}
\end{figure*}

Figure~\ref{fig:M_56_spec_scaled} shows synthetic spectra of the fiducial model with \sub{M}{56Ni}\ varied between $0.4-0.8$~\Msun.
Naturally, the bolometric luminosity increases proportionally with \sub{M}{56Ni}. The line ratios in these spectra are relatively insensitive to \sub{M}{56Ni}, with the exception being greater blending around the [FeIII] 5011 \AA\ feature in higher \fsix{Ni} mass models due to the larger size of the \fsix{Ni} core. The decrease in IME emission in higher \sub{M}{56Ni} models is due to the lower total mass of IMEs in these models by construction (since \sub{M}{56Ni} + \sub{M}{IME} is held fixed).

\subsection{Neutron-rich Iron Group Elements (IGEs)}
\label{ssec:M_IGE}

Stable IGEs are produced in SN~Ia ejecta in two distinct ways. The trace presence of neutron-rich isotopes (in particular $^{22}$Ne) due to the metallicity of the progenitor WD and pre-explosion carbon simmering leads to the production of up to $\approx 25$\% by mass of neutronized IGEs (in particular $^{54}$Fe and $^{58}$Ni) throughout the nickel core \citep{timmes2003, piro2008, martinez2016}. In addition, electron capture occurring in nuclear burning at high central densities can lead to the production of $\sim 0.05 - 0.4$~\Msun\ of neutronized IGEs \citep{nomoto1984, thielemann1986, seitenzahl2011, seitenzahl2013}. This latter effect only occurs in white dwarfs with $\rm M_{ej} \gtrsim M_{ch}$ (due to their higher central densities) and is also influenced by the timing of a possible deflagration-to-detonation transition \citep{seitenzahl2013}. 1D \sub{M}{Ch} models often predict stable IGEs to be produced at the core \citep{nomoto1984, mazzali2007a_hydro}, while multi-D simulations indicate that buoyancy should mix IGEs throughout the \fsix{Ni} region \citep{gamezo2005, kasen2009, seitenzahl2011}. As a result, nebular spectra indicators of stable IGEs would be valuable for inferring the progenitor scenario.

\begin{figure*}[htbp]
\includegraphics[width=0.99\textwidth]{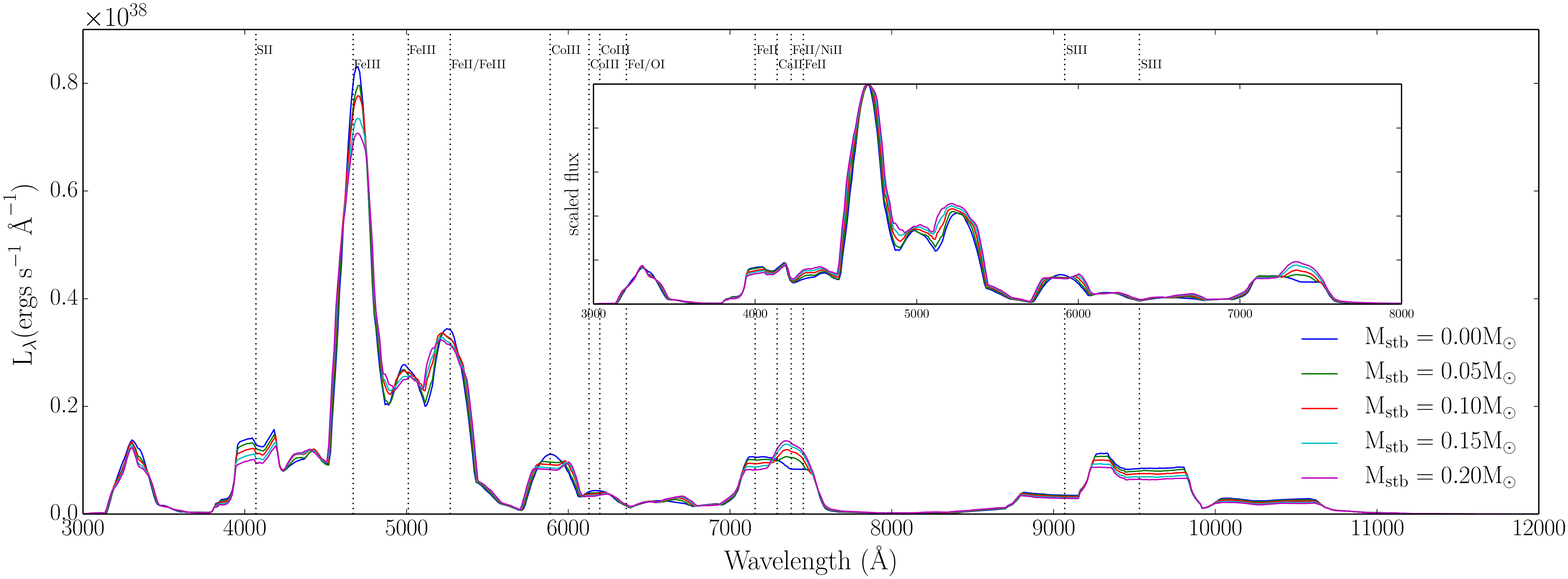}
\label{fig:M_IGE_spec_scaled}
\caption{Same as Figure \ref{fig:t_spec_scaled} but for the \sub{M}{stb} parameter study in which we add a region of stable IGEs of mass \sub{M}{stb} (and equal parts $^{58}$Ni and $^{54}$Fe) to the central core. We find that as little as 0.025 \Msun\ of central $^{58}$Ni is enough to produce detectable emission at 7378 \AA. The ``hole'' in energy deposition at the core also creates a characteristic flat-top profile for the [CoIII] 5888 \AA\ feature. }
\end{figure*}

Figure \ref{fig:M_IGE_spec_scaled} shows synthetic spectra of the fiducial model varying \sub{M}{stb}, the mass of a neutron-rich core, between $0.05 - 0.20$~\Msun. The stable IGEs are assumed to be an equal mix of $^{54}$Fe and $^{58}$Ni. As expected, the [NiII] 7378 \AA\ feature becomes apparent with a stable core mass of 0.05~\Msun, which corresponds to 0.025~\Msun\ of $^{58}$Ni.
[FeII] 7388 \AA\ emission can also produce a peak near these wavelengths, in some cases dominating the feature, suggesting that [NiII] may not even be needed to fit this peak (see \S\ref{sec:density_profiles}).

Higher stable core mass also increases the relative fraction of FeII compared to FeIII, which is a result of lowered non-thermal ionization in the non-radioactive core region. Another noteworthy product of a stable core is the flat-topped profile of the [CoIII] 5888 \AA\ feature, produced by the absence of Cobalt in the core region. The lower flux in IME features is due to the lower mass of IMEs in higher \sub{M}{stb} models by construction.

\begin{figure}[ht]
\includegraphics[width=0.5\textwidth]{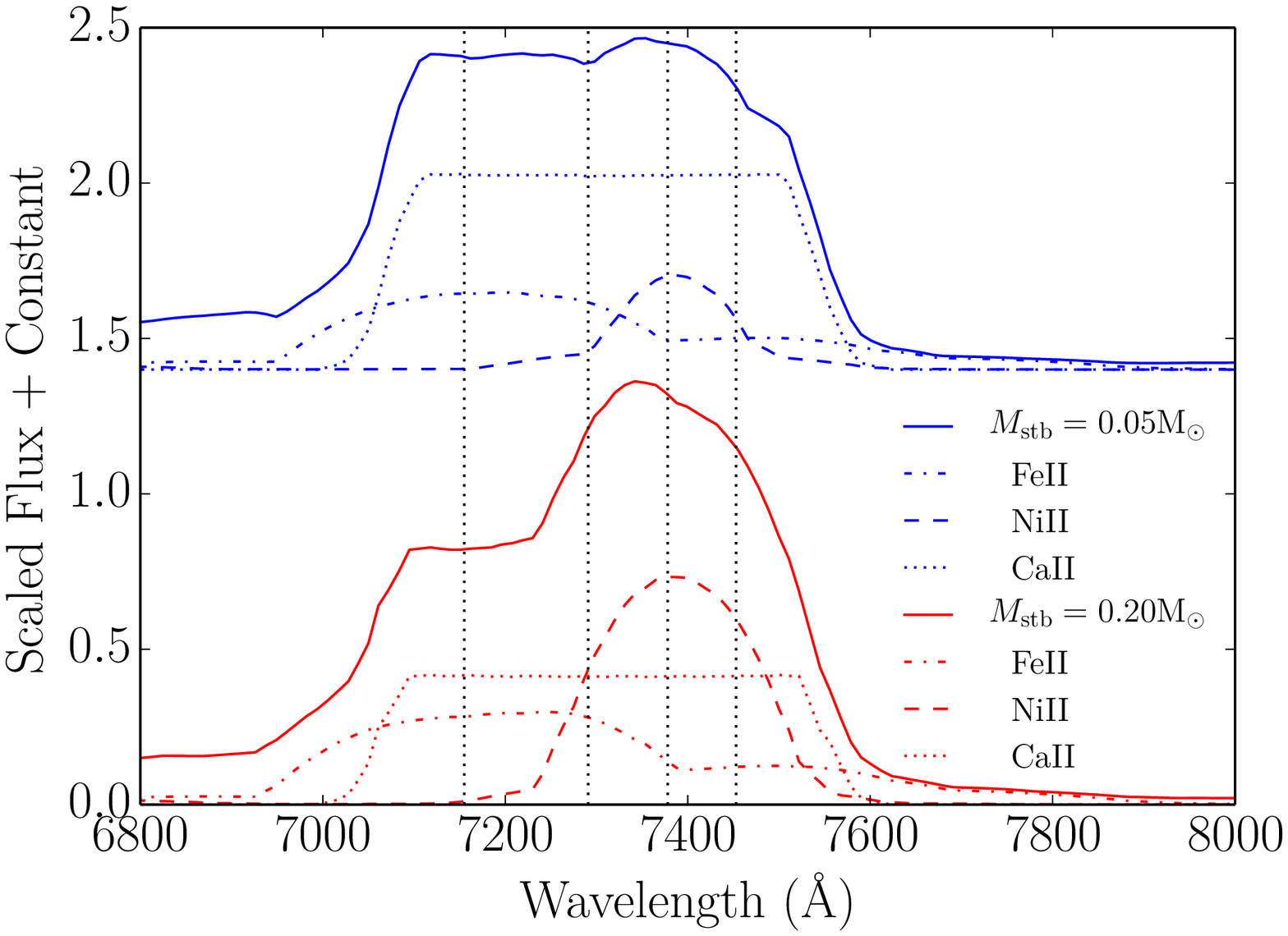}
\label{fig:M_IGE_NiII}
\caption{Decomposition of the spectral feature near [NiII]7388 \AA\ for  two models with varying stable core masses. The apparent blueshift (by $\approx 1500$~km/s) of the [NiII] peak in the $M_{\rm stb} = 0.2 M_\odot$ model is due to the relative blending between FeII and NiII components. The feature is also sensitive to [CaII] emission, which is stronger in the low-\sub{M}{stb} model. Vertical dotted lines show the line centers of the following transitions: [FeII] 7155 \AA, [CaII] 7291 \AA, [NiII] 7378 \AA, [FeII] 7388 \AA.  }

\end{figure}

A comparison between two \sub{M}{stb} models is shown in Figure \ref{fig:M_IGE_NiII}. In the $M_{\rm stb} = 0.2 M_\odot$ model we find an apparent blueshift of the [NiII] 7378 \AA\ peak by  $\approx 1500$ km/s due to the relative blending between FeII and NiII components. The feature is also sensitive to [CaII] emission, which is stronger in the low-\sub{M}{stb}\ model. The [FeII] feature redward of [NiII] 7378 \AA\ may also dominate in some models, resulting in an apparent redshift of the peak. 

\cite{maeda2010b_nature} find shifts in the 7378 \AA\ peak of up to 3000 km/s, which they interpret as indications of ejecta asymmetry. Our models suggest that this geometrical interpretation is complicated by line blending. Isolating the [NiII] emission in this feature can been attempted \citep{maeda2010a_asymmetry} but is subject to model-dependent uncertainties.

Surprisingly, the mixing of stable Nickel through the nickel zone (i.e., varying the parameter \sub{X}{stb}\ between $0.05 - 0.20$) does not produce a visible effect on the synthetic spectrum, even for $^{58}$Ni masses comparable to those in the core stable IGE region of figure \ref{fig:M_IGE_spec_scaled}. This is due to the high level of ionization of Ni in the nickel zone, which suppresses [NiII] emission. For the same reason, the ratio of $^{54}$Fe to $^{58}$Ni (\sub{R}{stb}) in the nickel zone also had no visible effect on the synthetic spectra in the range (0.5-1.5) that we tested. We predict that up to 0.1~\Msun\ of stable nickel can be hidden in the ejecta of a SN~Ia model with fiducial parameters due to this ionization effect. 

\subsection{Mass of Carbon/Oxygen in ejecta}
\label{ssec:M_CO}

Observational studies have estimated that about 30\% of SNe~Ia show carbon in their early-time spectra \citep[and references therein]{maoz2014}. The nearby SN2011fe also showed both carbon and oxygen in very early observations \citep{nugent2011, mazzali2014}, and hydrodynamical simulations have predicted various amounts of unburned C/O material mixed throughout the ejecta \citep{ropke2005, pakmor2012, seitenzahl2013, moll2014}. 3D nebular modeling by \citet{kozma2005} showed clear [OI] features for a pure deflagration model of \citet{ropke2005}, which contained 0.6\Msun\ of unburned C/O material. There have also been detections of possible [OI] emission in sub-luminous SNe~Ia \citep{taubenberger2013_oxygen_in_SnIa, kromer2013}.

\begin{figure*}[htbp]
\includegraphics[width=0.99\textwidth]{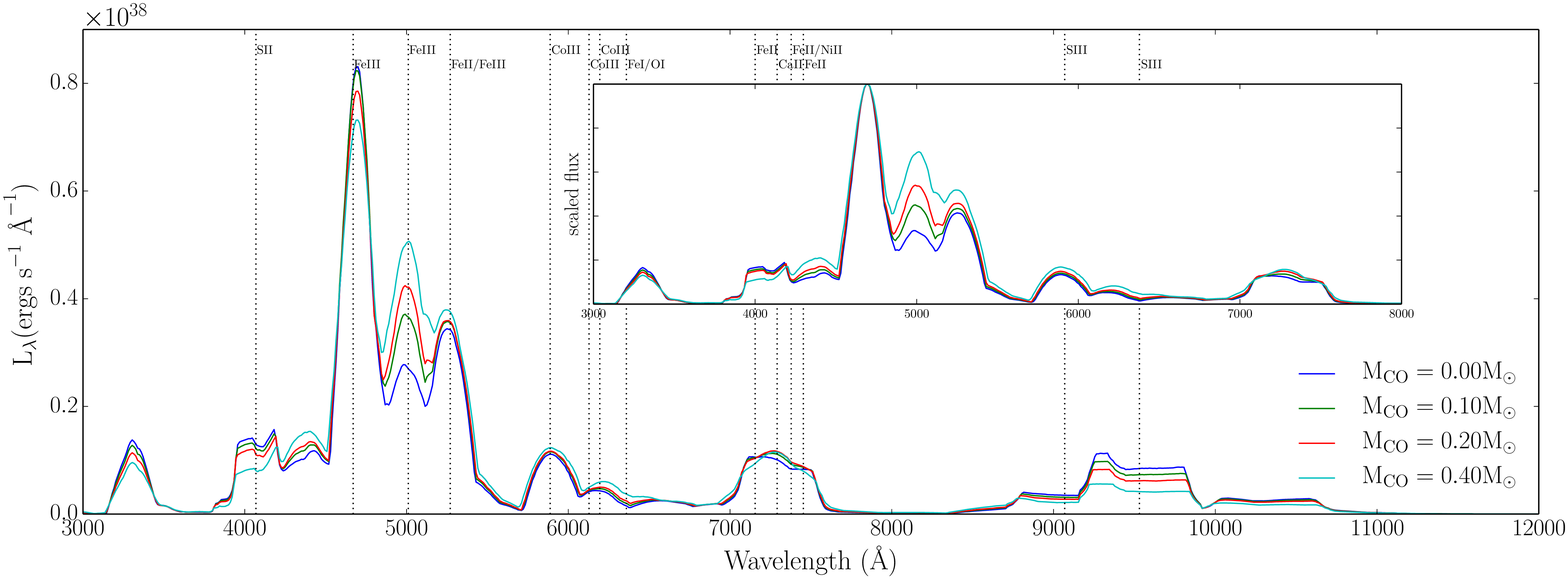}
\label{fig:M_CO_spec}
\caption{Same as Figure \ref{fig:t_spec_scaled} but for the \sub{M}{CO}\ parameter study in which we mix C/O into both nickel zone and IME layer (with a 1:9 carbon to oxygen ratio). Note the emergence of the [OIII] 5007 \AA\ feature with increasing C/O mass.}
\end{figure*}

Figure \ref{fig:M_CO_spec} shows synthetic spectra of the fiducial model with C/O mass varied between $0.1 - 0.4$~\Msun. We keep a fixed Carbon-Oxygen ratio of 1:9 and mix C/O into both the nickel zone and IME layer with the same mass fraction, consistent with the expected nucleosynthetic yields of delayed detonation explosions \citet{seitenzahl2013}.

We find a strong contribution of [OIII] at 5007 \AA\ as well a weak contribution of [OI] 6300/6364 \AA\ and a blend of [OII] features at 7320 \AA. The high ionization state of oxygen prevents the [OI] emission seen in \citet{kozma2005} to contribute significantly, but we expect [OI] to become stronger if oxygen were more concentrated in the higher density central regions, or if the mass of \fsix{Ni} (and hence radioactive deposition) were lower. C/O within the IME layer does not produce a significant nebular feature.

\subsection{Sensitivity to density profile}
\label{sec:density_profiles}

The choice of the ejecta density profile is a critical input into nebular modeling. A broken power-law  with a relatively flat interior was found to approximate the density structure in some 2D delayed detonation models \citep{kasen2010}, while a steeper exponential profile more closely fits the structure of the commonly-used 1D W7 model \citep{nomoto1984, thielemann1986}. Here, we attempt to illuminate the effect of steepening our interior power-law density profiles. We also consider an exponential profile of the form $\rm \rho = \rho_0 e^{-v/v_e}$ where $\rho_0$ and \sub{v}{e}\ can be determined from \sub{M}{ej}, \sub{E}{K}, and \sub{t}{ex}.

\begin{figure}[ht]
\includegraphics[width=0.5\textwidth]{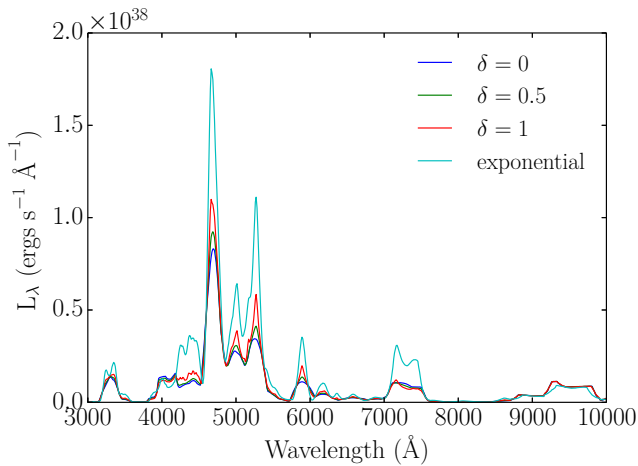}
\label{fig:dens_profile_comp}
\caption{Fiducial model with varying density profiles. $\delta$ refers to the exponent of the inner region in which $\rm \rho \propto v^{-\delta}$, and the exponential model has a density $\rm \rho \propto e^{-v/v_e}$ with a characteristic e-folding velocity \sub{v}{e}(\sub{M}{ej}, \sub{E}{K}). Steeper density profiles concentrate more mass towards the center and have narrower features, higher bolometric luminosities and lower ionization states (as indicated by the ratio of fluxes at 4665 \AA\ and 5272 \AA).}
\end{figure}

Figure \ref{fig:dens_profile_comp} shows synthetic spectra of the fiducial model with varied density profiles. Steeper profiles, which concentrate more mass at low velocities, produce stronger and narrower spectral profiles. In particular, an exponential profile helps resolve the individual peaks in the feature around 7300 \AA. The two peaks in this model are dominated by [FeII] and [CaII], with little contribution from [NiII].

Higher gamma-ray trapping due to higher central densities produces higher bolometric luminosities in steeper density profiles. Furthermore, steeper density profiles lead to lower central temperatures, as cooling becomes more efficient at higher densities. Therefore, the ionization state is lower, which changes the main FeIII/FeII line ratio.

When comparing to the nebular spectrum of SN2011fe, we find that a broken power-law with a flat interior profile ($\delta = 0$) best reproduces the shape of the features and the main iron line ratios, whereas an exponential density produces lines that are too centrally peaked and overestimates the FeII to FeIII ratio (indicated by high flux in the [FeII] feature at 5159 \AA). While there is some degeneracy with other parameters, such as kinetic energy, nebular models may be able constrain the interior ejecta density, which should be useful in testing explosion models.

\subsection{Sensitivity to atomic data uncertainties}
\label{ssec:atomic_study}

One important factor for the modeling of nebular spectra is the extensive atomic data inputs. Uncertainties in published atomic data, discrepancies between different sources for the same data, and the need for crude approximations where data are lacking impact the model predictions. Here we attempt to quantify some of the uncertainties by systematically changing the values of three of the most important and uncertain atomic data -- the collisional ionization cross sections ($\rm Q_k$), the collisional excitation rates ($\rm C_{ij}$), and the radiative recombination rates ($\alpha$).

\begin{figure}[ht]
\includegraphics[width=0.5\textwidth]{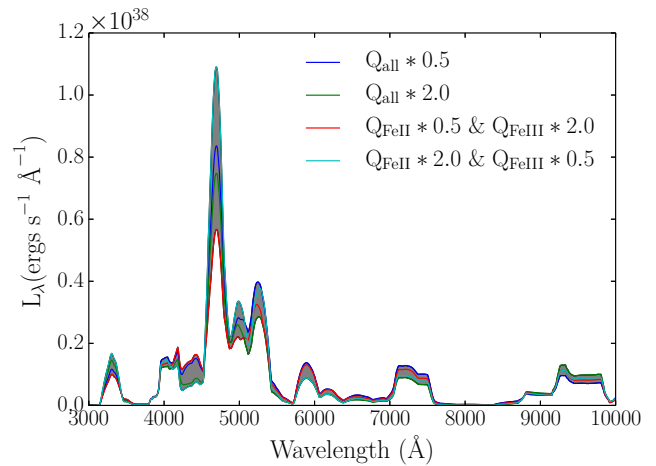}
\label{fig:fid_IC_noise}
\caption{Fiducial model in which collisional ionization cross section values were scaled by the same factor. Uncertainties at the factor of 2 level in the cross-sections produce $\approx 25\%$ changes in the spectral features.
}
\end{figure}

\begin{figure}[ht]
\includegraphics[width=0.5\textwidth]{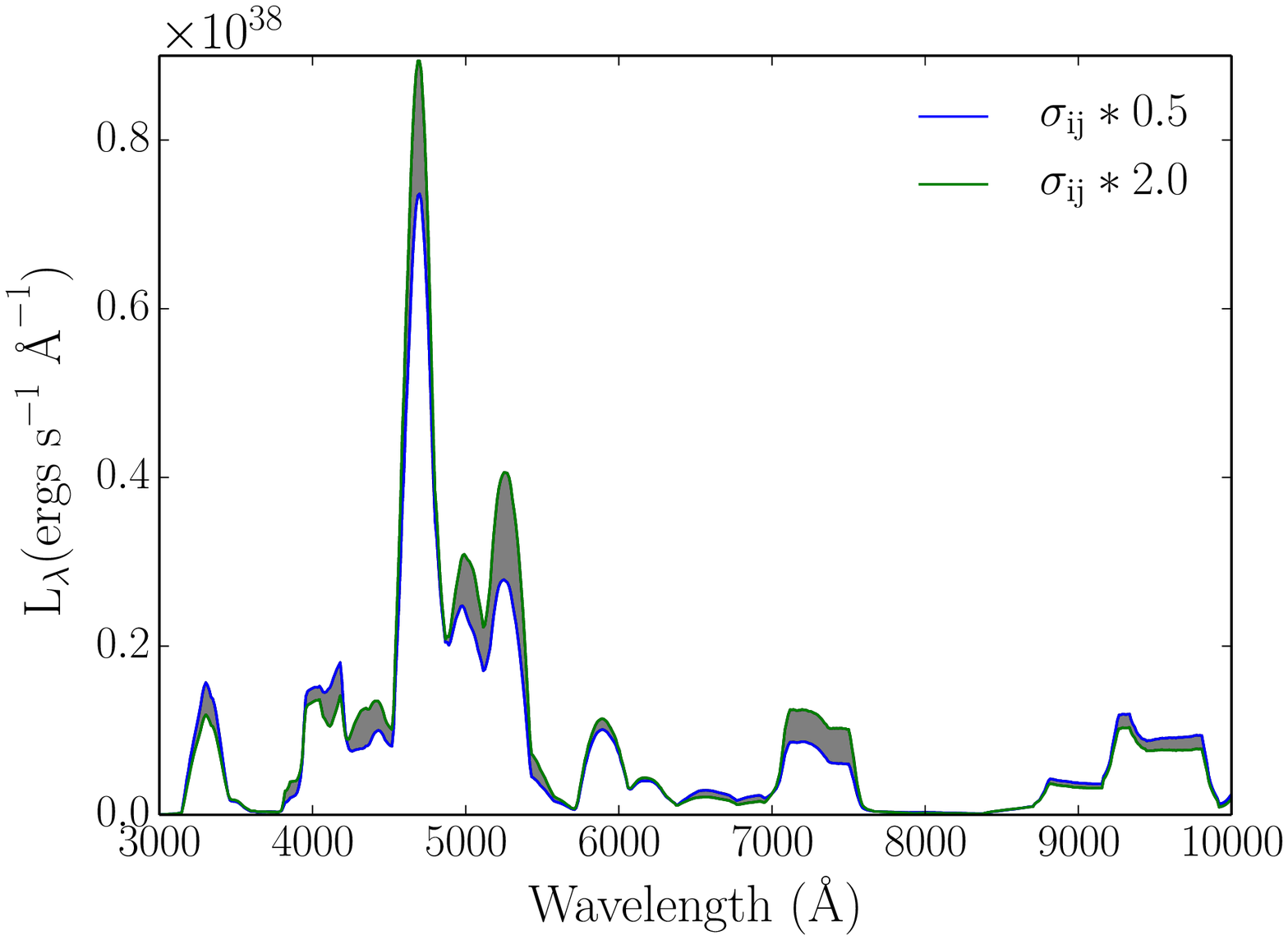}
\label{fig:fid_IJ_noise}
\caption{Fiducial model in which collisional excitation rates for all transitions of all species where scaled by the same factor.
}
\end{figure}

We show the impact of systematically varying atomic data in Figures \ref{fig:fid_IC_noise}-\ref{fig:fid_IJ_noise}. To explore the effect of the collisional ionization cross sections  (Figure~\ref{fig:fid_IC_noise}), we carried out calculations in which (a) all ionization cross sections were increased by a factor of 2, (b) all ionization cross sections were decreased by a factor of 2, (c) the cross section of FeII was decreased by a factor of 2 and that of FeIII was increased by a factor of 2, and (d) the cross section of FeII was increased by a factor of 2 and that of FeIII was decreased by a factor of 2. These variations in $Q$ affect the ionization ratio of FeIII/FeII and can modify the FeIII/FeII line ratio by up to $\pm$25\%. There are also indirect effects, since changes to the atomic data of one element can alter the calculated gas temperature and so result in changes in the emission from other species. 

To explore the effect of the collisional excitation rates  (Figure~\ref{fig:fid_IJ_noise}), we carried out calculations in which all \sub{\sigma}{ij} were increased or decreased by a factor of 2. These variations in \sub{\sigma}{ij} affect the strength of emission features and can modify the FeIII/FeII line ratio by up to $\pm$10\%.

\begin{figure}[ht]
\includegraphics[width=0.5\textwidth]{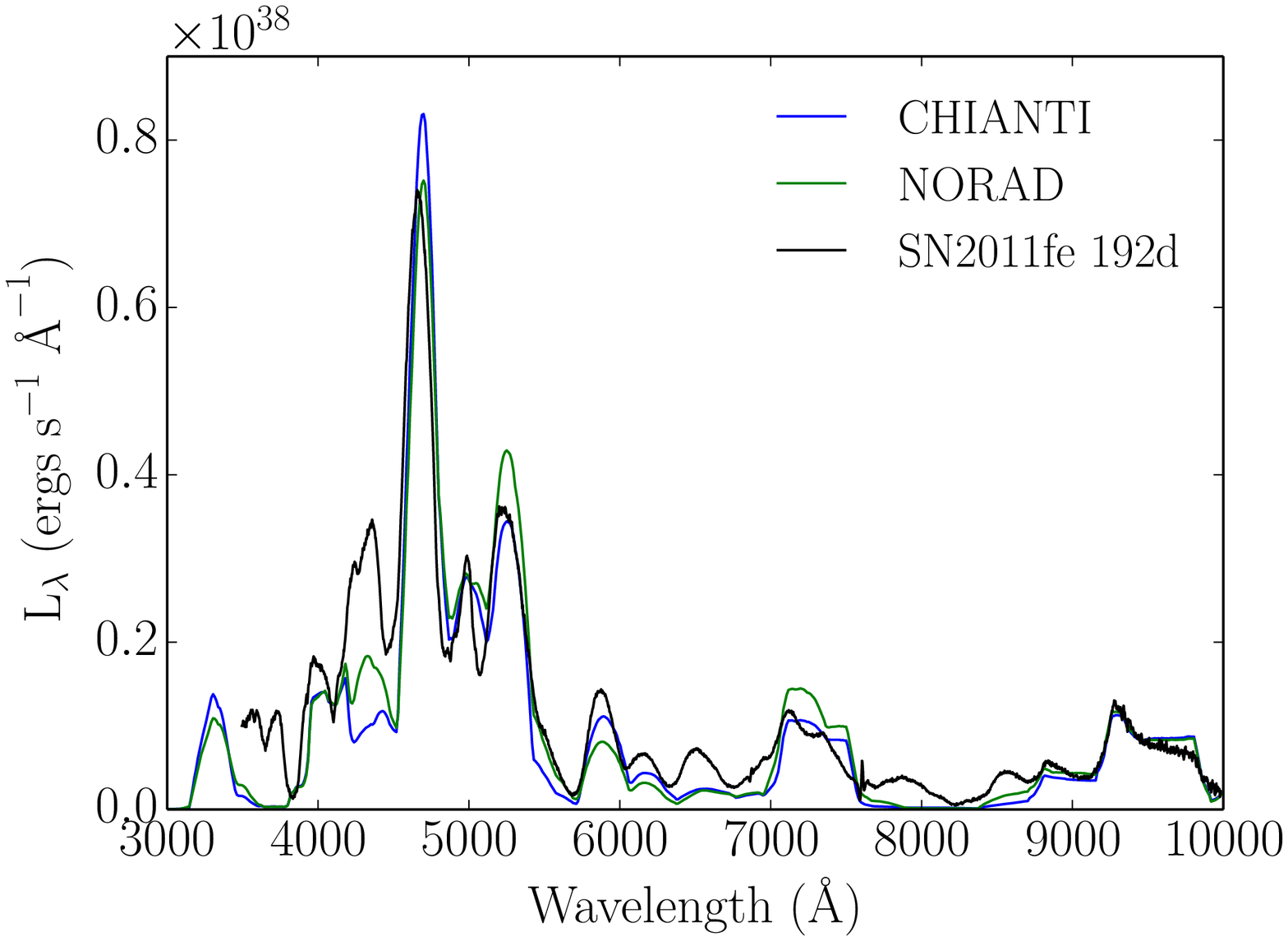}
\label{fig:norad_alpha_comp}
\caption{Fiducial model with two different sets of recombination rates for FeI-V.}
\end{figure}

We  show in Figure~\ref{fig:norad_alpha_comp} the fiducial model calculated using recombination rates from two different databases. Our calculations throughout has used recombination rates from the CHIANTI database, but more recent data for FeI-V are available from the Nahar OSU Radiative Atomic Database (NORAD) \citep{nahar1996_FeIII, nahar1997_FeI, nahar1997_FeII, nahar1998_FeIV}. With this latter data set, we have access to state-specific recombination rates, which we neglected in the above treatment. We also neglected charge transfer in the above analyses, which becomes important for a nickel zone with low ionization state (as is the case with NORAD recombination rates). We therefore include charge transfer in FeI-IV with data from \citet{krstic1997}.

While there is reasonable agreement between the synthetic spectra, the strength of emission features does depend on the atomic data set. In particular, the FeIII/FeII line ratio shows a higher FeII population when NORAD recombination rates are used, due to higher total recombination rates for FeI-III in the NORAD data set, leading to a $\sim$30\% decrease in FeIII/FeII line ratio.
Similar variations are seen in most other spectral features.

\section{DISCUSSION and conclusion}
\label{sec:conclusion}

We have presented a new tool for calculating nebular spectra of SNe and applied it to a systematic parameter study of SNe~Ia. We summarize some of our main results as follows:

\newpoint{Robustness and Degeneracy of Spectral Features} On the whole, we found that the features in SN~Ia nebular spectra were remarkably insensitive when physical parameters were changed within the range expected for white dwarf explosions. Individually changing the ejected mass, \fsix{Ni} mass, or kinetic energy by 50\% produced only minor changes in the relative strength of features (although the bolometric luminosity was affected due to changes in gamma-ray trapping). Varying parameters by a larger factor of $\approx 2$ began to show noticeable spectral changes. 

In addition, we found degeneracies in the effect of different physical parameters. An increase in the total ejecta mass, for example, could be mostly offset by a corresponding increase in kinetic energy so as to keep the overall density and velocity scale roughly fixed. As a result, quite different sets of physical parameters may be able to fit the same nebular spectrum.

\newpoint{Progenitor Mass} Though some previous studies have favored Chandrasekhar mass models in explaining SN~Ia nebular spectra, our model survey demonstrates that observed nebular spectra are equally well fit by generic sub-Chandrasekhar mass, Chandrasekhar mass, and super-Chandrasekhar mass models. The spectral features remained essentially unchanged when the total mass, \fsix{Ni} mass, and kinetic energy were scaled up or down in unison (Figure~\ref{fig:M_scl_spec}). Nebular spectra therefore do not alone constrain the overall mass scale of SNe~Ia, and instead are more diagnostic of the relative abundance yields and the density profile. 

\newpoint{Stable Iron Group Elements} Previous studies have fit the nebular spectra of SNe~Ia with models having a $\sim 0.1-0.2~M_\odot$ sphere of stable IGEs at the ejecta center. The presence of such an IGE ``core" would favor a Chandrasekhar mass model (where electron capture occurs during high-density deflagration burning), while disfavoring double-detonation and violent merger models. 

A stable IGE ``core" has been claimed necessary to explain the FeIII/FeII line ratio in observed spectra \citep{maeda2010a_asymmetry, mazzali2015}. We show that this line ratio also depends on other explosion parameters, such as the density profile and kinetic energy, and is sensitive to the atomic data inputs. We present models that well fit the SN~2011fe iron line ratio well without invoking any stable IGE ``core."

Our models that did include a large ($\gtrsim 0.1~M_\odot$) stable IGE ``core" produced a flat-topped [CoIII] feature near 5900 \AA. Such a flat-top [CoIII] profile is not seen in the nebular spectra of SN~2011fe \citep{mazzali2015} indicating that at least some radioactive cobalt exists at the lowest ejecta velocities. \citet{hoflich2004} and \citet{motohara2006} observe relatively flat-topped profiles of the 1.64~$\mu$m [FeII] feature in SN~2003du and SN~2003hv, respectively, which they take as evidence that SNe~Ia have a stable IGE ``core" lacking any radioactive heating. The lack of an obvious flat-topped optical [CoIII] feature in SN~2011fe appears to provide evidence to the contrary. 

Another diagnostic of stable IGE is the feature at $\sim 7300$ \AA\ usually attributed to [NiII]. Similar to  \citep{mazzali2015}, we find that a small amount ($\sim 0.01~M_\odot$) of central $^{58}$Ni is sufficient to reproduce a peak at 7300 \AA. However, we note that in some models the [FeII] 7388 \AA\ line can alone account for this peak, while in general [FeII] 7155 \AA\ and [CaII] emission also shape the overall line profile. The steepness of the inner ejecta density profile also affects how narrow and resolved the separate line peaks of the 7300 \AA\ feature are. We find that additional $^{58}$Ni mixed throughout the lower density layers of radioactive \fsix{Ni} zone may be too ionized to produce significant [NiII] 7378 \AA\ emission. As a result, we consider it difficult derive a precise $^{58}$Ni mass constraint  from analysis of the 7300 \AA\ feature. 

In sum, our parameter study does not provide strong evidence that SNe~Ia possess a substantial stable IGE ``core," nor do we see a robust way of accurately inferring the total stable IGE mass from nebular spectrum analysis. We therefore do not consider the nebular spectra of SNe~Ia as providing particularly strong support for the Chandrasekhar mass model over other progenitor scenarios. However, our study did not focus on infrared features \citep{hoflich2004, motohara2006, gerardy2007, telesco2015} or the effects of positron transport \citep{penney2014}. Further study of stable IGE signatures using specific multi-dimensional explosion models is warranted.

\newpoint{IME indicators} While the nebular spectra of SNe~Ia are dominated by iron lines, we emphasize that some features provide  constraints on IME abundances. In particular, the feature at 7300 \AA\ is sensitive to [CaII] emission at 7291 \AA, while the feature at 9500 \AA\ is dominated by [SIII] emission. We showed that the relative strength of the [SIII] 9500 \AA\ feature to the [FeIII] 4664 \AA\ feature tracked the relative abundance of IGE to IME (see Figure \ref{fig:M_56_spec_scaled}). Analysis of this line ratio may thus provide valuable constraints on the nucleosynthetic yields and hence progenitors of SNe~Ia. Our successful fit to the calcium and sulfur features in SN~2011fe suggests that the total IME yields may be $\approx 0.75~M_\odot$.

\newpoint{Line Shifts and Ejecta Geometry} Shifts in the wavelength of nebular line peaks are seen in many SNe~Ia \citep{black2016} and have been used to deduce asymmetries of, e.g., iron group elements \citep{maeda2010a_asymmetry}. Though we have only considered spherically symmetric models in this paper, we nevertheless see shifts in the location of emission peaks due to line blending. For example, the feature around 7300 \AA\ is a blend of [NiII], [FeII], and [CaII] lines. Depending on the ejecta compositional structure and time-evolution of line strengths, the location of the composite peak can vary by 1500 km/s in different models.
This highlights the difficulty in using the 7300~\AA\ feature shift to derive reliable kinematic measures of asymmetry (though see \cite{maeda2010a_asymmetry} for attempts separate components in the blend). Our models do not show a strong, progressive redshift evolution in the main IGE features as a function of time. This supports the claim of \cite{black2016} that these shifts are due to permitted line emission that may help shape the spectrum even at times $\gtrsim 200$~days after explosion.

\newpoint{Carbon/Oxygen mixing} We find that C/O material mixed into a nickel zone is highly ionized. For large enough C/O masses ($\gtrsim 0.1 M_\odot)$ this produces a visible [OIII] feature at 5007 \AA, which is not seen in SN~2011fe. This disfavors models leaving significant amounts of oxygen mixed throughout the ejecta. To produce a narrow [OI] feature like that seen by \citet{taubenberger2013_oxygen_in_SnIa} in the sub-luminous SN~2010lp
would presumably require the C/O to to be located in a dense central region with less non-thermal ionization from radioactivity. C/O in the outermost ejecta layers (within or above the IME zone) does not experience significant radioactive heating and has no impact on the nebular spectra.

\newpoint{Density Profile} We have shown that ejecta having a steep interior density profile produce narrower and more central peaked nebular emission features. In fact, the line widths depended more on the density profile than on the kinetic energy, since the former sets the degree of central concentration of mass. Comparing to the spectra of SN~2011fe, we find that a flat density profile (power-law index $\delta = 0 - 1$) best reproduces the observed spectral profiles. This suggests that nebular spectra may be able to test the density structures predicted by detailed SN~Ia explosion models.

\newpoint{Atomic Data} Though it is well known that synthetic nebular spectra are affected by uncertainty and incompleteness in the input atomic data, the level of error has not been well quantified. We showed here that factor of 2 errors in some of the key atomic data inputs can result in changes in spectral features at the $\approx 25\%$ level. This level of variation is similar to the level we found when varying the physical ejecta parameters. Given the uncertainties, we advise against over-interpreting model fits to extract quantitative mass estimates from individual observed SNe~Ia. The trends seen among a sample of nebular spectra, however, are likely less affected by the atomic data limitations.

\newpoint{Limitations} While the spectral models presented here included a broad range of the most important and complex atomic processes, other physical effects may be relevant and should be addressed in future work. While the ejecta are mostly optically thin at optical wavelengths, there is evidence that some lines are nevertheless optically thick, especially in the ultraviolet \citep{friesen2017}. Furthermore, dust/molecule absorption and emission may be important at some phases. At very late times, the freezing out of ionization requires that the problem be treated time-dependently \citep{kozma1998a,jerkstrand2015}. The collection of a reliable and complete atomic data sets has been and remains a substantial challenge for all nebular phase modeling efforts. 

Finally, while this paper was restricted to 1D parameterized models, the asphericity found in more realistic explosion models, and inferred from polarization observation of SNe, is expected to affect the nebular spectra. Our NLTE nebular spectrum code is 3D, and future work on SNe~Ia and other types of SNe will aim to explore the predictions of 3D hydrodynamical models that have complex abundance distributions, clumping on multiple scales, and global ejecta asymmetries. 

\medskip

\acknowledgments{
  This material is based upon work supported by the National Science Foundation Graduate Research Fellowship Program under Grant No. DGE 1106400. DK is supported in part by a Department of Energy Office of Nuclear Physics Early Career Award, and by the Director, Office of Energy Research, Office of High Energy and Nuclear Physics, Divisions of Nuclear Physics, of the U.S. Department of Energy under Contract No. DE-AC02-05CH11231.
}
\appendix
\section{Atomic Data} \label{appx:atomicData}

We use the atomic data compilation of the \textsc{cmfgen} code \citep{hillier1998, li2012} for radiative transition probabilities (Einstein A Coefficients), effective collisional strengths $\Upsilon_{\rm ij}$ (see appendix \ref{ssec:C_ij} for treatment of missing data), and photoionization cross sections. We take ground-state ionization energies from NIST \citep{kramida2016_NIST}. For thermal collisional ionization rates, we use the results of \citet{voronov1997}. For collisional ionization cross sections, we use \citet{mattioli2007} and references therein. Radiative and di-electronic recombination rates are taken from the CHIANTI database v.8 \citep{dere1997,delzanna2015}. 

\section{Nonthermal Deposition Fractions}
\label{appx:nonThermal}

Following the Continuous Slowing Down Approximation formulation of \citet{xu1991}, we can write the fraction of deposition into the three channels heating ($\eta_{\rm h}$), ionization ($\eta_{\rm k}$), and excitation ($\eta_{ij}$) as such

\begin{equation}
\label{eqn:eta_h}
\eta_{\rm h} =  \frac{L_{\rm e}(E_0) + f\sum_{\rm k} x_{\rm k} I_{\rm k} Q_{\rm k}(E_0)}{A}
\end{equation}

\begin{equation}
\eta_{\rm k} = \frac{x_{\rm k} I_{\rm k} Q_{\rm k}(E_0)}{A}
\end{equation}
where $E_0$ is the injection energy of electrons, $L_e$ is the Coulomb loss function (see below), k labels an ion and its ionization potential $I_{\rm k}$, ionization fraction $x_{\rm k} = n_{\rm k}/n$, and collisional ionization cross section $Q_{\rm k}(E)$, and $f \sim 0.3$ is the mean energy of a secondary electron, and

\begin{equation}
\eta_{\rm ij} = \frac{x_{\rm i} E_{\rm ij} \sigma_{\rm ij}(E_0)}{A}
\end{equation}
where ij labels a transition with energy $E_{\rm ij} = E_{\rm j} - E_{\rm i} > 0$, the fractional level population of the lower-energy state is $x_{\rm i} = n_{\rm i} / n$, and the collisional excitation cross section is $\sigma_{\rm ij}$, and 

\begin{equation}
\label{eqn:A_nt}
A =  L_{\rm e}(E_0) + (1+f)\sum_{\rm k} x_{\rm k} I_{\rm k} Q_{\rm k}(E_0) + \sum_{\rm ij} x_{\rm ij} E_{\rm ij} \sigma_{\rm ij}(E_0)
\end{equation}
is the normalization constant. The Coulomb loss function \citep{schunk1971, xu1991} for high-energy electrons is

\begin{equation}
L_{\rm e}(E) = -\frac{1}{n}\left(\frac{dE}{dx}\right)_{\rm e} = x_{\rm e} \frac{2\pi e^4}{E}\ln\left(\frac{4 E}{\zeta_{\rm e}}\right)
\end{equation}
where $E$ is the electron energy, $e$ is the electron charge, $x_{\rm e} = n_{\rm e}/n$ is the electron fraction, and $\zeta_{\rm e} = 2 \hbar \omega_{\rm p}$ for plasma frequency given by 

\begin{equation}
\omega_{\rm p}(n_{\rm e}) = \sqrt{\frac{4 \pi n_{\rm e} e^2}{m_{\rm e}}} = 56414.6 \sqrt{\frac{n_{\rm e}}{\rm cm^{-3}}}~\rm s^{-1}
\end{equation}

\noindent For collisional ionization cross sections, we use the approximate form \citep{younger1981}
\begin{equation}
Q_{\rm I}(E) = \frac{1}{uI^2}\left[A\left(1 - \frac{1}{u}\right)+B\left(
1-\frac{1}{u}\right)^2  + C~\ln(u) + D \frac{\ln(u)}{u}\right]~\rm cm^2
\end{equation}
where $u = E/I$, I is the ionization potential, and A, B, C, and D are fitting coefficients. We use the fitting coefficients published by \citet{arnaud1985}, taking into consideration updates by \citet{arnaud1992} and \citet{mattioli2007}. While more recent calculations are available, \citet{dere2007} points out that the discrepancies between new and old ionization rates are minor except for NiV-XI, which are not important for this work. Coefficients are unavailable for CoII-IV, so we use FeII-IV cross sections as instead and plan to update this data when it becomes available in the future. 

For collisional excitation cross sections of allowed transitions by non-thermal electrons, we use the approximation of \citet{vanReg1962},
\begin{equation}
\sigma_{\rm ij} = \frac{8 \pi}{\sqrt{3}} \frac{1}{k_{\rm i}^2} \frac{I_{\rm H}}{\Delta E_{\rm ij}} f_{\rm ij}\bar{g}\pi a_0^2
\end{equation}
where $I_{\rm H}$ is the ionization potential of Hydrogen, $k_{\rm i}^2$ is the initial electron energy scaled to 13.60 eV, $f_{\rm ij}$ is the oscillator strength, $a_0$ is the Bohr radius, and $\bar{g}$ is given by a quadratic fit to the results of \citet{vanReg1962}

\begin{equation}
\bar{g} \sim  -0.0065\left(\frac{E}{\Delta E_{\rm ij}}\right)^2 + 0.228\left(\frac{E}{\Delta E_{\rm ij}}\right)-0.07 
\end{equation}
with an imposed high-energy limit of 
\begin{equation}
\bar{g} = \rm \frac{\sqrt{3}}{2\pi}ln\left(\frac{E}{\Delta E_{ij}}\right) , \ E/\Delta E_{ij} > 36
\end{equation}
for all ions, and a low-energy limit for positive ions

\begin{equation}
\bar{g} = \rm 0.2, \ E / \Delta E_{ij} < \sqrt{2}
\end{equation}

\noindent Forbidden transitions must be treated separately. We use the following form for the collision cross sections: 
\begin{equation}
\sigma_{\rm ij} =  \frac{1}{g_{\rm i}k_{\rm i}^2} \Omega_{\rm ij}\bar{g}\pi a_0^2
\end{equation}
where $g_i$ is the statistical weight and $\Omega_{ij}$ is the collision strength given by equation \ref{equ:van_reg}.

\section{Implementation of Atomic Processes}
\label{appx:implementation}

\subsection{Collisional excitation and de-excitation}
\label{ssec:C_ij}

To determine collisional excitation and de-excitation rates, we assume electrons adhere to a Maxwell-Boltzmann distribution. The subsequent ``effective'' or ``Maxwellian-averaged'' collisional strength can be calculated for each transition using

\begin{equation}
	\label{equ:upsilon}
   \Upsilon_{\rm ij} = \int_0^\infty \Omega_{\rm ij}e^{-\epsilon_{\rm j}/kT} d(\epsilon_{\rm j}/kT)
\end{equation}
where $\Omega{\rm ij}$ is the collisional cross section for an electron to excite the i-j transition. These collision strengths can be found in the literature and data tables.

The subsequent rates of collisional excitation and de-excitation, respectively, are

\begin{eqnarray}
	\label{equ:C_ij}
	C_{\rm ij}(T) &=& \frac{8.63 \times 10^{-6}}{g_{\rm i}\sqrt{T}}e^{-E_{\rm ij}/kT} \Upsilon_{\rm ij} n_{\rm e}~\rm s^{-1} \T \\
	C_{\rm ji}(T) &=& \frac{g_{\rm i}}{g_{\rm j}} e^{E_{\rm ij} / k_{\rm B}T}  C_{\rm ij} \T
\end{eqnarray}
where $E_{ij}$ is the energy difference between the two levels and $j > i$.

In cases where atomic data for $\Upsilon$ are unavailable, one can use the dipole-approximation \citep{vanReg1962, jefferies1968}, which gives results for neutral atoms and positive ions, respectively,

\begin{equation}
	\label{equ:van_reg}
	C_{\rm ij}(T)  =
    \begin{cases}
    \displaystyle 2.16 f_{\rm ij} n_{\rm e}  \biggl[ \frac{E_{\rm ij}}{kT} \biggr]^{-1.68} T^{-3/2} e^{-E_{\rm ij}/kT} \rm \ s^{-1}, ~(neutral~atoms) \T \\
	 \displaystyle 3.9 f_{\rm ij} n_{\rm e}  \biggl[ \frac{E_{\rm ij}}{kT} \biggr]^{-1} T^{-3/2} e^{-E_{\rm ij}/kT}~ \rm s^{-1},~(positive~ions) \T
    \end{cases}
\end{equation}
where $f_{\rm ij}$ is oscillator strength, $T$ is Temperature in Kelvin, $E_{\rm ij}$ is transition energy in eV, and $n_{\rm e}$ is electron density in $\rm cm^{-3}$. 

In practice, atomic data for $\Upsilon_{\rm ij}$ are often published in a narrow temperature range; for temperatures outside this range, we use the above approximation to fill in the missing data. We scale equation \ref{equ:van_reg} so that there is no discontinuity in $C_{\rm ij}$.

Forbidden transitions for which atomic data is not available should be treated as a special case, since oscillator strengths for these transitions produce artificially low collisional rates. \citet{axelrod1980} provides the following formulation for forbidden transitions (defined as $f_{\rm ij} \leq 0.001$):

\begin{equation}
\label{equ:omega_axelrod}
\Omega_{\rm ij, forb} = 
\begin{cases}
0.00375 ~g_{\rm i}g_{\rm j}, ~\lambda \leq 10\mu \\
0.0225 ~g_{\rm i}g_{\rm j}, ~\lambda > 10\mu
\end{cases}
\end{equation}

\subsection{Recombination}
Total recombination rates of relevant ions include a sum over all sub-shells and incorporate both radiative and dielectronic recombination. We assume recombinations involve only ground states, a good approximation in regimes where collisional excitation dominates the population of excited states.  We use the rates published in the CHIANTI database v.8 \citep{dere1997, delzanna2015}. Although level-specific recombination rates are available for certain ions \citep[etc.]{nahar1992, nahar1994}, we opt to use a consistent source for recombination rates for this work. We explore this latter dataset in \S\ref{ssec:atomic_study} and plan to implement state-specific recombinations rates in any future work where recombination lines might contribute significantly. 

\medskip
\bibliography{biblio}

\begin{thebibliography}{}
\expandafter\ifx\csname natexlab\endcsname\relax\def\natexlab#1{#1}\fi

\bibitem[{{Arnaud} \& {Raymond}(1992)}]{arnaud1992}
{Arnaud}, M., \& {Raymond}, J. 1992, \apj, 398, 394

\bibitem[{{Arnaud} \& {Rothenflug}(1985)}]{arnaud1985}
{Arnaud}, M., \& {Rothenflug}, R. 1985, \aaps, 60, 425

\bibitem[{{Ashall} {et~al.}(2016){Ashall}, {Mazzali}, {Pian}, \&
  {James}}]{ashall2016}
{Ashall}, C., {Mazzali}, P.~A., {Pian}, E., \& {James}, P.~A. 2016, \mnras,
  463, 1891

\bibitem[{{Axelrod}(1980)}]{axelrod1980}
{Axelrod}, T.~S. 1980, PhD thesis, California Univ., Santa Cruz.

\bibitem[{{Benz} {et~al.}(1989){Benz}, {Thielemann}, \& {Hills}}]{benz1989}
{Benz}, W., {Thielemann}, F.-K., \& {Hills}, J.~G. 1989, \apj, 342, 986

\bibitem[{{Black} {et~al.}(2016){Black}, {Fesen}, \& {Parrent}}]{black2016}
{Black}, C.~S., {Fesen}, R.~A., \& {Parrent}, J.~T. 2016, \mnras, 462, 649

\bibitem[{{Branch} {et~al.}(1993){Branch}, {Fisher}, \& {Nugent}}]{branch1993}
{Branch}, D., {Fisher}, A., \& {Nugent}, P. 1993, \aj, 106, 2383

\bibitem[{{Branch} \& {Khokhlov}(1995)}]{branch1995}
{Branch}, D., \& {Khokhlov}, A.~M. 1995, \physrep, 256, 53

\bibitem[{{Bravo} {et~al.}(2009){Bravo}, {Garc{\'{\i}}a-Senz}, {Cabez{\'o}n},
  \& {Dom{\'{\i}}nguez}}]{bravo2009b}
{Bravo}, E., {Garc{\'{\i}}a-Senz}, D., {Cabez{\'o}n}, R.~M., \&
  {Dom{\'{\i}}nguez}, I. 2009, \apj, 695, 1257

\bibitem[{{Chevalier} \& {Soker}(1989)}]{chevalier1989}
{Chevalier}, R.~A., \& {Soker}, N. 1989, \apj, 341, 867

\bibitem[{{Colgate} \& {McKee}(1969)}]{colgate1969}
{Colgate}, S.~A., \& {McKee}, C. 1969, \apj, 157, 623

\bibitem[{{Del Zanna} {et~al.}(2015){Del Zanna}, {Dere}, {Young}, {Landi}, \&
  {Mason}}]{delzanna2015}
{Del Zanna}, G., {Dere}, K.~P., {Young}, P.~R., {Landi}, E., \& {Mason}, H.~E.
  2015, \aap, 582, A56

\bibitem[{{Dere}(2007)}]{dere2007}
{Dere}, K.~P. 2007, \aap, 466, 771

\bibitem[{{Dere} {et~al.}(1997){Dere}, {Landi}, {Mason}, {Monsignori Fossi}, \&
  {Young}}]{dere1997}
{Dere}, K.~P., {Landi}, E., {Mason}, H.~E., {Monsignori Fossi}, B.~C., \&
  {Young}, P.~R. 1997, \aaps, 125, doi:10.1051/aas:1997368

\bibitem[{{Dessart} \& {Hillier}(2011)}]{dessart2011}
{Dessart}, L., \& {Hillier}, D.~J. 2011, \mnras, 410, 1739

\bibitem[{{Dimitriadis} {et~al.}(2017){Dimitriadis}, {Sullivan}, {Kerzendorf},
  {Ruiter}, {Seitenzahl}, {Taubenberger}, {Doran}, {Gal-Yam}, {Laher},
  {Maguire}, {Nugent}, {Ofek}, \& {Surace}}]{dimitriadis2017}
{Dimitriadis}, G., {Sullivan}, M., {Kerzendorf}, W., {et~al.} 2017, ArXiv
  e-prints, arXiv:1701.07267

\bibitem[{{Fink} {et~al.}(2007){Fink}, {Hillebrandt}, \&
  {R{\"o}pke}}]{fink2007}
{Fink}, M., {Hillebrandt}, W., \& {R{\"o}pke}, F.~K. 2007, \aap, 476, 1133

\bibitem[{{Fransson} \& {Jerkstrand}(2015)}]{fransson2015}
{Fransson}, C., \& {Jerkstrand}, A. 2015, \apjl, 814, L2

\bibitem[{{Fransson} \& {Kozma}(1993)}]{fransson1993}
{Fransson}, C., \& {Kozma}, C. 1993, \apjl, 408, L25

\bibitem[{{Friesen} {et~al.}(2017){Friesen}, {Baron}, {Parrent}, {Thomas},
  {Branch}, {Nugent}, {Hauschildt}, {Foley}, {Wright}, {Pan}, {Filippenko},
  {Clubb}, {Silverman}, {Maeda}, {Shivvers}, {Kelly}, {Cohen}, {Rest}, \&
  {Kasen}}]{friesen2017}
{Friesen}, B., {Baron}, E., {Parrent}, J.~T., {et~al.} 2017, \mnras,
  arXiv:1607.04784

\bibitem[{{Gamezo} {et~al.}(2005){Gamezo}, {Khokhlov}, \& {Oran}}]{gamezo2005}
{Gamezo}, V.~N., {Khokhlov}, A.~M., \& {Oran}, E.~S. 2005, \apj, 623, 337

\bibitem[{{Gerardy} {et~al.}(2007){Gerardy}, {Meikle}, {Kotak}, {H{\"o}flich},
  {Farrah}, {Filippenko}, {Foley}, {Lundqvist}, {Mattila}, {Pozzo},
  {Sollerman}, {Van Dyk}, \& {Wheeler}}]{gerardy2007}
{Gerardy}, C.~L., {Meikle}, W.~P.~S., {Kotak}, R., {et~al.} 2007, \apj, 661,
  995

\bibitem[{{Gnat} \& {Ferland}(2012)}]{gnat2012}
{Gnat}, O., \& {Ferland}, G.~J. 2012, \apjs, 199, 20

\bibitem[{{Golombek} \& {Niemeyer}(2005)}]{golombek2005}
{Golombek}, I., \& {Niemeyer}, J.~C. 2005, \aap, 438, 611

\bibitem[{{Guillochon} {et~al.}(2010){Guillochon}, {Dan}, {Ramirez-Ruiz}, \&
  {Rosswog}}]{guillochon2010}
{Guillochon}, J., {Dan}, M., {Ramirez-Ruiz}, E., \& {Rosswog}, S. 2010, \apjl,
  709, L64

\bibitem[{{Hillebrandt} {et~al.}(2013){Hillebrandt}, {Kromer}, {R{\"o}pke}, \&
  {Ruiter}}]{hillebrandt2013}
{Hillebrandt}, W., {Kromer}, M., {R{\"o}pke}, F.~K., \& {Ruiter}, A.~J. 2013,
  Frontiers of Physics, 8, 116

\bibitem[{{Hillier} \& {Miller}(1998)}]{hillier1998}
{Hillier}, D.~J., \& {Miller}, D.~L. 1998, \apj, 496, 407

\bibitem[{{H{\"o}flich} {et~al.}(2004){H{\"o}flich}, {Gerardy}, {Nomoto},
  {Motohara}, {Fesen}, {Maeda}, {Ohkubo}, \& {Tominaga}}]{hoflich2004}
{H{\"o}flich}, P., {Gerardy}, C.~L., {Nomoto}, K., {et~al.} 2004, \apj, 617,
  1258

\bibitem[{{Howell}(2011)}]{howell2011}
{Howell}, D.~A. 2011, Nature Communications, 2, 350

\bibitem[{{Hoyle} \& {Fowler}(1960)}]{hoyle1960}
{Hoyle}, F., \& {Fowler}, W.~A. 1960, \apj, 132, 565

\bibitem[{{Iben} \& {Tutukov}(1984)}]{iben1984}
{Iben}, Jr., I., \& {Tutukov}, A.~V. 1984, \apjs, 54, 335

\bibitem[{{Iwamoto} {et~al.}(1999){Iwamoto}, {Brachwitz}, {Nomoto},
  {Kishimoto}, {Umeda}, {Hix}, \& {Thielemann}}]{iwamoto1999}
{Iwamoto}, K., {Brachwitz}, F., {Nomoto}, K., {et~al.} 1999, \apjs, 125, 439

\bibitem[{{Jefferies}(1968)}]{jefferies1968}
{Jefferies}, J.~T. 1968, {Spectral line formation}

\bibitem[{{Jerkstrand} {et~al.}(2015){Jerkstrand}, {Ergon}, {Smartt},
  {Fransson}, {Sollerman}, {Taubenberger}, {Bersten}, \&
  {Spyromilio}}]{jerkstrand2015}
{Jerkstrand}, A., {Ergon}, M., {Smartt}, S.~J., {et~al.} 2015, \aap, 573, A12

\bibitem[{{Jerkstrand} {et~al.}(2011){Jerkstrand}, {Fransson}, \&
  {Kozma}}]{jerkstrand2011_1987A}
{Jerkstrand}, A., {Fransson}, C., \& {Kozma}, C. 2011, \aap, 530, A45

\bibitem[{{Jordan} {et~al.}(2008){Jordan}, {Fisher}, {Townsley}, {Calder},
  {Graziani}, {Asida}, {Lamb}, \& {Truran}}]{jordan2008}
{Jordan}, IV, G.~C., {Fisher}, R.~T., {Townsley}, D.~M., {et~al.} 2008, \apj,
  681, 1448

\bibitem[{{Jordan} {et~al.}(2012){Jordan}, {Graziani}, {Fisher}, {Townsley},
  {Meakin}, {Weide}, {Reid}, {Norris}, {Hudson}, \& {Lamb}}]{jordan2012}
{Jordan}, IV, G.~C., {Graziani}, C., {Fisher}, R.~T., {et~al.} 2012, \apj, 759,
  53

\bibitem[{{Kasen}(2010)}]{kasen2010}
{Kasen}, D. 2010, \apj, 708, 1025

\bibitem[{{Kasen} {et~al.}(2009){Kasen}, {R{\"o}pke}, \& {Woosley}}]{kasen2009}
{Kasen}, D., {R{\"o}pke}, F.~K., \& {Woosley}, S.~E. 2009, \nat, 460, 869

\bibitem[{{Kasen} {et~al.}(2006){Kasen}, {Thomas}, \& {Nugent}}]{kasen2006}
{Kasen}, D., {Thomas}, R.~C., \& {Nugent}, P. 2006, \apj, 651, 366

\bibitem[{{Kashyap} {et~al.}(2015){Kashyap}, {Fisher}, {Garc{\'{\i}}a-Berro},
  {Aznar-Sigu{\'a}n}, {Ji}, \& {Lor{\'e}n-Aguilar}}]{kashyep2015}
{Kashyap}, R., {Fisher}, R., {Garc{\'{\i}}a-Berro}, E., {et~al.} 2015, \apjl,
  800, L7

\bibitem[{{Kerzendorf} {et~al.}(2014){Kerzendorf}, {Taubenberger},
  {Seitenzahl}, \& {Ruiter}}]{kerzendorf2014}
{Kerzendorf}, W.~E., {Taubenberger}, S., {Seitenzahl}, I.~R., \& {Ruiter},
  A.~J. 2014, \apjl, 796, L26

\bibitem[{{Kozma} \& {Fransson}(1998{\natexlab{a}})}]{kozma1998a}
{Kozma}, C., \& {Fransson}, C. 1998{\natexlab{a}}, \apj, 496, 946

\bibitem[{{Kozma} \& {Fransson}(1998{\natexlab{b}})}]{kozma1998b}
---. 1998{\natexlab{b}}, \apj, 497, 431

\bibitem[{{Kozma} {et~al.}(2005){Kozma}, {Fransson}, {Hillebrandt},
  {Travaglio}, {Sollerman}, {Reinecke}, {R{\"o}pke}, \&
  {Spyromilio}}]{kozma2005}
{Kozma}, C., {Fransson}, C., {Hillebrandt}, W., {et~al.} 2005, \aap, 437, 983

\bibitem[{{Kramida} {et~al.}(2016){Kramida}, {Ralchenko}, \&
  {Reader}}]{kramida2016_NIST}
{Kramida}, A., {Ralchenko}, Y., \& {Reader}, J. 2016, in APS Division of
  Atomic, Molecular and Optical Physics Meeting Abstracts

\bibitem[{{Kromer} {et~al.}(2013){Kromer}, {Pakmor}, {Taubenberger}, {Pignata},
  {Fink}, {R{\"o}pke}, {Seitenzahl}, {Sim}, \& {Hillebrandt}}]{kromer2013}
{Kromer}, M., {Pakmor}, R., {Taubenberger}, S., {et~al.} 2013, \apjl, 778, L18

\bibitem[{{Krstic} {et~al.}(1997){Krstic}, {Stancil}, \&
  {Schultz}}]{krstic1997}
{Krstic}, P.~S., {Stancil}, P.~C., \& {Schultz}, D.~R. 1997, NASA STI/Recon
  Technical Report N, 99

\bibitem[{{Kuchner} {et~al.}(1994){Kuchner}, {Kirshner}, {Pinto}, \&
  {Leibundgut}}]{kuchner1994}
{Kuchner}, M.~J., {Kirshner}, R.~P., {Pinto}, P.~A., \& {Leibundgut}, B. 1994,
  \apjl, 426, 89

\bibitem[{{Kushnir} {et~al.}(2013){Kushnir}, {Katz}, {Dong}, {Livne}, \&
  {Fern{\'a}ndez}}]{kushnir2013}
{Kushnir}, D., {Katz}, B., {Dong}, S., {Livne}, E., \& {Fern{\'a}ndez}, R.
  2013, \apjl, 778, L37

\bibitem[{{Leloudas} {et~al.}(2009){Leloudas}, {Stritzinger}, {Sollerman},
  {Burns}, {Kozma}, {Krisciunas}, {Maund}, {Milne}, {Filippenko}, {Fransson},
  {Ganeshalingam}, {Hamuy}, {Li}, {Phillips}, {Schmidt}, {Skottfelt},
  {Taubenberger}, {Boldt}, {Fynbo}, {Gonzalez}, {Salvo}, \&
  {Thomas-Osip}}]{leloudas2009}
{Leloudas}, G., {Stritzinger}, M.~D., {Sollerman}, J., {et~al.} 2009, \aap,
  505, 265

\bibitem[{{Li} {et~al.}(2012){Li}, {Hillier}, \& {Dessart}}]{li2012}
{Li}, C., {Hillier}, D.~J., \& {Dessart}, L. 2012, \mnras, 426, 1671

\bibitem[{{Li} \& {McCray}(1993)}]{liMcCray1993}
{Li}, H., \& {McCray}, R. 1993, \apj, 405, 730

\bibitem[{{Liu} {et~al.}(1997){Liu}, {Jeffery}, \&
  {Schultz}}]{liu1997a_nebular}
{Liu}, W., {Jeffery}, D.~J., \& {Schultz}, D.~R. 1997, \apjl, 483, L107

\bibitem[{{Livne}(1990)}]{livne1990a_double_detonation}
{Livne}, E. 1990, \apjl, 354, L53

\bibitem[{{Maeda} {et~al.}(2006){Maeda}, {Nomoto}, {Mazzali}, \&
  {Deng}}]{maeda2006_TypeIc_multidim}
{Maeda}, K., {Nomoto}, K., {Mazzali}, P.~A., \& {Deng}, J. 2006, \apj, 640, 854

\bibitem[{{Maeda} {et~al.}(2010{\natexlab{a}}){Maeda}, {Taubenberger},
  {Sollerman}, {Mazzali}, {Leloudas}, {Nomoto}, \&
  {Motohara}}]{maeda2010a_asymmetry}
{Maeda}, K., {Taubenberger}, S., {Sollerman}, J., {et~al.} 2010{\natexlab{a}},
  \apj, 708, 1703

\bibitem[{{Maeda} {et~al.}(2010{\natexlab{b}}){Maeda}, {Benetti},
  {Stritzinger}, {R{\"o}pke}, {Folatelli}, {Sollerman}, {Taubenberger},
  {Nomoto}, {Leloudas}, {Hamuy}, {Tanaka}, {Mazzali}, \&
  {Elias-Rosa}}]{maeda2010b_nature}
{Maeda}, K., {Benetti}, S., {Stritzinger}, M., {et~al.} 2010{\natexlab{b}},
  \nat, 466, 82

\bibitem[{{Maoz} {et~al.}(2014){Maoz}, {Mannucci}, \& {Nelemans}}]{maoz2014}
{Maoz}, D., {Mannucci}, F., \& {Nelemans}, G. 2014, \araa, 52, 107

\bibitem[{{Marietta} {et~al.}(2000){Marietta}, {Burrows}, \&
  {Fryxell}}]{marietta2000}
{Marietta}, E., {Burrows}, A., \& {Fryxell}, B. 2000, \apjs, 128, 615

\bibitem[{{Mart{\'{\i}}nez-Rodr{\'{\i}}guez}
  {et~al.}(2016){Mart{\'{\i}}nez-Rodr{\'{\i}}guez}, {Piro}, {Schwab}, \&
  {Badenes}}]{martinez2016}
{Mart{\'{\i}}nez-Rodr{\'{\i}}guez}, H., {Piro}, A.~L., {Schwab}, J., \&
  {Badenes}, C. 2016, \apj, 825, 57

\bibitem[{{Mattioli} {et~al.}(2007){Mattioli}, {Mazzitelli}, {Finkenthal},
  {Mazzotta}, {Fournier}, {Kaastra}, \& {Puiatti}}]{mattioli2007}
{Mattioli}, M., {Mazzitelli}, G., {Finkenthal}, M., {et~al.} 2007, Journal of
  Physics B Atomic Molecular Physics, 40, 3569

\bibitem[{{Maurer} {et~al.}(2011){Maurer}, {Jerkstrand}, {Mazzali},
  {Taubenberger}, {Hachinger}, {Kromer}, {Sim}, \& {Hillebrandt}}]{maurer2011}
{Maurer}, I., {Jerkstrand}, A., {Mazzali}, P.~A., {et~al.} 2011, \mnras, 418,
  1517

\bibitem[{{Maurer} {et~al.}(2010){Maurer}, {Mazzali}, {Taubenberger}, \&
  {Hachinger}}]{maurer2010}
{Maurer}, I., {Mazzali}, P.~A., {Taubenberger}, S., \& {Hachinger}, S. 2010,
  \mnras, 409, 1441

\bibitem[{{Mazzali} \& {Hachinger}(2012)}]{mazzali2012}
{Mazzali}, P.~A., \& {Hachinger}, S. 2012, \mnras, 424, 2926

\bibitem[{{Mazzali} {et~al.}(2011){Mazzali}, {Maurer}, {Stritzinger},
  {Taubenberger}, {Benetti}, \& {Hachinger}}]{mazzali2011}
{Mazzali}, P.~A., {Maurer}, I., {Stritzinger}, M., {et~al.} 2011, \mnras, 416,
  881

\bibitem[{{Mazzali} {et~al.}(2001){Mazzali}, {Nomoto}, {Patat}, \&
  {Maeda}}]{mazzali2001}
{Mazzali}, P.~A., {Nomoto}, K., {Patat}, F., \& {Maeda}, K. 2001, \apj, 559,
  1047

\bibitem[{{Mazzali} {et~al.}(2007{\natexlab{a}}){Mazzali}, {R{\"o}pke},
  {Benetti}, \& {Hillebrandt}}]{mazzali2007a_hydro}
{Mazzali}, P.~A., {R{\"o}pke}, F.~K., {Benetti}, S., \& {Hillebrandt}, W.
  2007{\natexlab{a}}, Science, 315, 825

\bibitem[{{Mazzali} {et~al.}(2007{\natexlab{b}}){Mazzali}, {Kawabata}, {Maeda},
  {Foley}, {Nomoto}, {Deng}, {Suzuki}, {Iye}, {Kashikawa}, {Ohyama},
  {Filippenko}, {Qiu}, \& {Wei}}]{mazzali2007b_nebular}
{Mazzali}, P.~A., {Kawabata}, K.~S., {Maeda}, K., {et~al.} 2007{\natexlab{b}},
  \apj, 670, 592

\bibitem[{{Mazzali} {et~al.}(2014){Mazzali}, {Sullivan}, {Hachinger}, {Ellis},
  {Nugent}, {Howell}, {Gal-Yam}, {Maguire}, {Cooke}, {Thomas}, {Nomoto}, \&
  {Walker}}]{mazzali2014}
{Mazzali}, P.~A., {Sullivan}, M., {Hachinger}, S., {et~al.} 2014, \mnras, 439,
  1959

\bibitem[{{Mazzali} {et~al.}(2015){Mazzali}, {Sullivan}, {Filippenko},
  {Garnavich}, {Clubb}, {Maguire}, {Pan}, {Shappee}, {Silverman}, {Benetti},
  {Hachinger}, {Nomoto}, \& {Pian}}]{mazzali2015}
{Mazzali}, P.~A., {Sullivan}, M., {Filippenko}, A.~V., {et~al.} 2015, \mnras,
  450, 2631

\bibitem[{{Moll} {et~al.}(2014){Moll}, {Raskin}, {Kasen}, \&
  {Woosley}}]{moll2014}
{Moll}, R., {Raskin}, C., {Kasen}, D., \& {Woosley}, S.~E. 2014, \apj, 785, 105

\bibitem[{{Motohara} {et~al.}(2006){Motohara}, {Maeda}, {Gerardy}, {Nomoto},
  {Tanaka}, {Tominaga}, {Ohkubo}, {Mazzali}, {Fesen}, {H{\"o}flich}, \&
  {Wheeler}}]{motohara2006}
{Motohara}, K., {Maeda}, K., {Gerardy}, C.~L., {et~al.} 2006, \apjl, 652, L101

\bibitem[{{Nahar}(1996)}]{nahar1996_FeIII}
{Nahar}, S.~N. 1996, \pra, 53, 2417

\bibitem[{{Nahar}(1997)}]{nahar1997_FeII}
---. 1997, \pra, 55, 1980

\bibitem[{{Nahar} {et~al.}(1997){Nahar}, {Bautista}, \&
  {Pradhan}}]{nahar1997_FeI}
{Nahar}, S.~N., {Bautista}, M.~A., \& {Pradhan}, A.~K. 1997, \apj, 479, 497

\bibitem[{{Nahar} {et~al.}(1998){Nahar}, {Bautista}, \&
  {Pradhan}}]{nahar1998_FeIV}
---. 1998, \pra, 58, 4593

\bibitem[{{Nahar} \& {Pradhan}(1992)}]{nahar1992}
{Nahar}, S.~N., \& {Pradhan}, A.~K. 1992, Physical Review Letters, 68, 1488

\bibitem[{{Nahar} \& {Pradhan}(1994)}]{nahar1994}
---. 1994, \pra, 49, 1816

\bibitem[{{Nahar} \& {Pradhan}(1997)}]{nahar1997_C_N}
---. 1997, \apjs, 111, 339

\bibitem[{{Nomoto} {et~al.}(1984){Nomoto}, {Thielemann}, \&
  {Yokoi}}]{nomoto1984}
{Nomoto}, K., {Thielemann}, F.-K., \& {Yokoi}, K. 1984, \apj, 286, 644

\bibitem[{{Nozawa} {et~al.}(2011){Nozawa}, {Maeda}, {Kozasa}, {Tanaka},
  {Nomoto}, \& {Umeda}}]{nozawa2011}
{Nozawa}, T., {Maeda}, K., {Kozasa}, T., {et~al.} 2011, \apj, 736, 45

\bibitem[{{Nugent} {et~al.}(2011){Nugent}, {Sullivan}, {Cenko}, {Thomas},
  {Kasen}, {Howell}, {Bersier}, {Bloom}, {Kulkarni}, {Kandrashoff},
  {Filippenko}, {Silverman}, {Marcy}, {Howard}, {Isaacson}, {Maguire},
  {Suzuki}, {Tarlton}, {Pan}, {Bildsten}, {Fulton}, {Parrent}, {Sand},
  {Podsiadlowski}, {Bianco}, {Dilday}, {Graham}, {Lyman}, {James}, {Kasliwal},
  {Law}, {Quimby}, {Hook}, {Walker}, {Mazzali}, {Pian}, {Ofek}, {Gal-Yam}, \&
  {Poznanski}}]{nugent2011}
{Nugent}, P.~E., {Sullivan}, M., {Cenko}, S.~B., {et~al.} 2011, \nat, 480, 344

\bibitem[{{Pakmor} {et~al.}(2012){Pakmor}, {Kromer}, {Taubenberger}, {Sim},
  {R{\"o}pke}, \& {Hillebrandt}}]{pakmor2012}
{Pakmor}, R., {Kromer}, M., {Taubenberger}, S., {et~al.} 2012, \apjl, 747, L10

\bibitem[{{Penney} \& {Hoeflich}(2014)}]{penney2014}
{Penney}, R., \& {Hoeflich}, P. 2014, \apj, 795, 84

\bibitem[{{Phillips}(1993)}]{phillips1993}
{Phillips}, M.~M. 1993, \apjl, 413, L105

\bibitem[{{Piro} \& {Bildsten}(2008)}]{piro2008}
{Piro}, A.~L., \& {Bildsten}, L. 2008, \apj, 673, 1009

\bibitem[{{Plewa}(2007)}]{plewa2007}
{Plewa}, T. 2007, \apj, 657, 942

\bibitem[{{Raskin} {et~al.}(2009){Raskin}, {Timmes}, {Scannapieco}, {Diehl}, \&
  {Fryer}}]{raskin2009}
{Raskin}, C., {Timmes}, F.~X., {Scannapieco}, E., {Diehl}, S., \& {Fryer}, C.
  2009, \mnras, 399, L156

\bibitem[{{R{\"o}pke}(2005)}]{ropke2005}
{R{\"o}pke}, F.~K. 2005, \aap, 432, 969

\bibitem[{{R{\"o}pke} {et~al.}(2007){R{\"o}pke}, {Hillebrandt}, {Schmidt},
  {Niemeyer}, {Blinnikov}, \& {Mazzali}}]{ropke2007b}
{R{\"o}pke}, F.~K., {Hillebrandt}, W., {Schmidt}, W., {et~al.} 2007, \apj, 668,
  1132

\bibitem[{{R{\"o}pke} \& {Niemeyer}(2007)}]{ropke2007a}
{R{\"o}pke}, F.~K., \& {Niemeyer}, J.~C. 2007, \aap, 464, 683

\bibitem[{{Rosswog} {et~al.}(2009){Rosswog}, {Kasen}, {Guillochon}, \&
  {Ramirez-Ruiz}}]{rosswog2009}
{Rosswog}, S., {Kasen}, D., {Guillochon}, J., \& {Ramirez-Ruiz}, E. 2009,
  \apjl, 705, L128

\bibitem[{{Ruiz-Lapuente} \& {Lucy}(1992)}]{ruiz-Lapuente1992}
{Ruiz-Lapuente}, P., \& {Lucy}, L.~B. 1992, \apj, 400, 127

\bibitem[{{Scalzo} {et~al.}(2014{\natexlab{a}}){Scalzo}, {Aldering},
  {Antilogus}, {Aragon}, {Bailey}, {Baltay}, {Bongard}, {Buton},
  {Cellier-Holzem}, {Childress}, {Chotard}, {Copin}, {Fakhouri}, {Gangler},
  {Guy}, {Kim}, {Kowalski}, {Kromer}, {Nordin}, {Nugent}, {Paech}, {Pain},
  {Pecontal}, {Pereira}, {Perlmutter}, {Rabinowitz}, {Rigault}, {Runge},
  {Saunders}, {Sim}, {Smadja}, {Tao}, {Taubenberger}, {Thomas}, {Weaver}, \&
  {Nearby Supernova Factory}}]{scalzo2014a}
{Scalzo}, R., {Aldering}, G., {Antilogus}, P., {et~al.} 2014{\natexlab{a}},
  \mnras, 440, 1498

\bibitem[{{Scalzo} {et~al.}(2014{\natexlab{b}}){Scalzo}, {Ruiter}, \&
  {Sim}}]{scalzo2014b}
{Scalzo}, R.~A., {Ruiter}, A.~J., \& {Sim}, S.~A. 2014{\natexlab{b}}, \mnras,
  445, 2535

\bibitem[{{Scalzo} {et~al.}(2010){Scalzo}, {Aldering}, {Antilogus}, {Aragon},
  {Bailey}, {Baltay}, {Bongard}, {Buton}, {Childress}, {Chotard}, {Copin},
  {Fakhouri}, {Gal-Yam}, {Gangler}, {Hoyer}, {Kasliwal}, {Loken}, {Nugent},
  {Pain}, {P{\'e}contal}, {Pereira}, {Perlmutter}, {Rabinowitz}, {Rau},
  {Rigaudier}, {Runge}, {Smadja}, {Tao}, {Thomas}, {Weaver}, \&
  {Wu}}]{scalzo2010}
{Scalzo}, R.~A., {Aldering}, G., {Antilogus}, P., {et~al.} 2010, \apj, 713,
  1073

\bibitem[{{Schunk} \& {Hays}(1971)}]{schunk1971}
{Schunk}, R.~W., \& {Hays}, P.~B. 1971, \planss, 19, 113

\bibitem[{{Seitenzahl} {et~al.}(2011){Seitenzahl}, {Ciaraldi-Schoolmann}, \&
  {R{\"o}pke}}]{seitenzahl2011}
{Seitenzahl}, I.~R., {Ciaraldi-Schoolmann}, F., \& {R{\"o}pke}, F.~K. 2011,
  \mnras, 414, 2709

\bibitem[{{Seitenzahl} {et~al.}(2013){Seitenzahl}, {Ciaraldi-Schoolmann},
  {R{\"o}pke}, {Fink}, {Hillebrandt}, {Kromer}, {Pakmor}, {Ruiter}, {Sim}, \&
  {Taubenberger}}]{seitenzahl2013}
{Seitenzahl}, I.~R., {Ciaraldi-Schoolmann}, F., {R{\"o}pke}, F.~K., {et~al.}
  2013, \mnras, 429, 1156

\bibitem[{{Shappee} {et~al.}(2013){Shappee}, {Stanek}, {Pogge}, \&
  {Garnavich}}]{shappee2013}
{Shappee}, B.~J., {Stanek}, K.~Z., {Pogge}, R.~W., \& {Garnavich}, P.~M. 2013,
  \apjl, 762, L5

\bibitem[{{Shen} \& {Bildsten}(2014)}]{shen2014}
{Shen}, K.~J., \& {Bildsten}, L. 2014, \apj, 785, 61

\bibitem[{{Shivvers} {et~al.}(2013){Shivvers}, {Mazzali}, {Silverman},
  {Boty{\'a}nszki}, {Cenko}, {Filippenko}, {Kasen}, {Van Dyk}, \&
  {Clubb}}]{shivvers2013}
{Shivvers}, I., {Mazzali}, P., {Silverman}, J.~M., {et~al.} 2013, \mnras, 436,
  3614

\bibitem[{{Sollerman} {et~al.}(2000){Sollerman}, {Kozma}, {Fransson},
  {Leibundgut}, {Lundqvist}, {Ryde}, \& {Woudt}}]{sollerman2000}
{Sollerman}, J., {Kozma}, C., {Fransson}, C., {et~al.} 2000, \apjl, 537, L127

\bibitem[{{Sollerman} {et~al.}(2004){Sollerman}, {Lindahl}, {Kozma}, {Challis},
  {Filippenko}, {Fransson}, {Garnavich}, {Leibundgut}, {Li}, {Lundqvist},
  {Milne}, {Spyromilio}, \& {Kirshner}}]{sollerman2004}
{Sollerman}, J., {Lindahl}, J., {Kozma}, C., {et~al.} 2004, \aap, 428, 555

\bibitem[{{Stehle} {et~al.}(2005){Stehle}, {Mazzali}, {Benetti}, \&
  {Hillebrandt}}]{stehle2005}
{Stehle}, M., {Mazzali}, P.~A., {Benetti}, S., \& {Hillebrandt}, W. 2005,
  \mnras, 360, 1231

\bibitem[{{Stritzinger} {et~al.}(2006){Stritzinger}, {Leibundgut}, {Walch}, \&
  {Contardo}}]{stritzinger2006a}
{Stritzinger}, M., {Leibundgut}, B., {Walch}, S., \& {Contardo}, G. 2006, \aap,
  450, 241

\bibitem[{{Sutherland} \& {Dopita}(1993)}]{sutherland1993}
{Sutherland}, R.~S., \& {Dopita}, M.~A. 1993, \apjs, 88, 253

\bibitem[{{Swartz}(1994)}]{swartz1994}
{Swartz}, D.~A. 1994, \apj, 428, 267

\bibitem[{{Swartz} {et~al.}(1995){Swartz}, {Sutherland}, \&
  {Harkness}}]{swartz1995}
{Swartz}, D.~A., {Sutherland}, P.~G., \& {Harkness}, R.~P. 1995, \apj, 446, 766

\bibitem[{{Taubenberger} {et~al.}(2013){Taubenberger}, {Kromer}, {Pakmor},
  {Pignata}, {Maeda}, {Hachinger}, {Leibundgut}, \&
  {Hillebrandt}}]{taubenberger2013_oxygen_in_SnIa}
{Taubenberger}, S., {Kromer}, M., {Pakmor}, R., {et~al.} 2013, \apjl, 775, L43

\bibitem[{{Telesco} {et~al.}(2015){Telesco}, {H{\"o}flich}, {Li},
  {{\'A}lvarez}, {Wright}, {Barnes}, {Fern{\'a}ndez}, {Hough}, {Levenson},
  {Mari{\~n}as}, {Packham}, {Pantin}, {Rebolo}, {Roche}, \&
  {Zhang}}]{telesco2015}
{Telesco}, C.~M., {H{\"o}flich}, P., {Li}, D., {et~al.} 2015, \apj, 798, 93

\bibitem[{{Thielemann} {et~al.}(1986){Thielemann}, {Nomoto}, \&
  {Yokoi}}]{thielemann1986}
{Thielemann}, F.-K., {Nomoto}, K., \& {Yokoi}, K. 1986, \aap, 158, 17

\bibitem[{{Timmes} {et~al.}(2003){Timmes}, {Brown}, \& {Truran}}]{timmes2003}
{Timmes}, F.~X., {Brown}, E.~F., \& {Truran}, J.~W. 2003, \apjl, 590, L83

\bibitem[{{van Regemorter}(1962)}]{vanReg1962}
{van Regemorter}, H. 1962, \apj, 136, 906

\bibitem[{{Voronov}(1997)}]{voronov1997}
{Voronov}, G.~S. 1997, Atomic Data and Nuclear Data Tables, 65, 1

\bibitem[{{Wang} \& {Han}(2012)}]{wang2012}
{Wang}, B., \& {Han}, Z. 2012, \nar, 56, 122

\bibitem[{{Webbink}(1984)}]{webbink1984}
{Webbink}, R.~F. 1984, \apj, 277, 355

\bibitem[{{Whelan} \& {Iben}(1973)}]{whelan1973}
{Whelan}, J., \& {Iben}, Jr., I. 1973, \apj, 186, 1007

\bibitem[{{Woosley} {et~al.}(2007){Woosley}, {Kasen}, {Blinnikov}, \&
  {Sorokina}}]{woosley2007}
{Woosley}, S.~E., {Kasen}, D., {Blinnikov}, S., \& {Sorokina}, E. 2007, \apj,
  662, 487

\bibitem[{{Woosley} \& {Weaver}(1994)}]{woosley1994}
{Woosley}, S.~E., \& {Weaver}, T.~A. 1994, \apj, 423, 371

\bibitem[{{Xu} \& {McCray}(1991)}]{xu1991}
{Xu}, Y., \& {McCray}, R. 1991, \apj, 375, 190

\bibitem[{{Younger}(1981)}]{younger1981}
{Younger}, S.~M. 1981, \jqsrt, 26, 329

\end{thebibliography}
\end{document}